\documentclass[aps,prb,twocolumn,10pt,article,nofootinbib,showpacs]{revtex4-1}

\usepackage{amsmath}
\usepackage{amssymb}
\usepackage{amsthm}
\usepackage{amsfonts}
\usepackage{listings}
\usepackage{enumerate}
\usepackage{latexsym}
\usepackage{graphicx}
\usepackage[caption=false]{subfig}
\usepackage{blkarray}
\usepackage{array}
\usepackage{color}
\usepackage[normalem]{ulem}
\usepackage{hyperref}

\def\y{\'{\i}}
\def\to{\rightarrow}

\def\non{\nonumber }

\def\x{\frac{k_x}{2}}
\def\y{\frac{\sqrt{3}}{2} k_y}
\def\xx{k_x}
\def\yy{\sqrt{3} k_y}

\newcommand{\beq}{\begin{equation}} 
\newcommand{\eeq}{\end{equation}} 
\newcommand{\beqa}{\begin{eqnarray}} 
\newcommand{\eeqa}{\end{eqnarray}}

\makeatletter
\DeclareRobustCommand{\element}[1]{\@element#1\@nil}
\def\@element#1#2\@nil{%
  #1%
  \if\relax#2\relax\else\MakeLowercase{#2}\fi}
\makeatother

\begin{document}
\title{Fragile phonon topology on the honeycomb lattice with time-reversal symmetry}
\author{Juan L. Ma\~nes}
\affiliation{Departamento de F\'\i sica de la Materia Condensada, Universidad del Pa\'\i s Vasco UPV/EHU}

\date{\today}
\begin{abstract}

We use the methods of Topological Quantum Chemistry to explore the topology of phonons on  time-reversal symmetric crystals with the structure of the planar honeycomb (layer group $p6/mmm$). This approach is not tied to a particular model of atomic vibrations, but is applied  to the most general dynamical matrix constrained only by the symmetries of the system. We show that four distinct  fragile topological  phonon phases are generically possible. Truncating the dynamical matrix to  third nearest neighbors yields a model that realizes the different phonon topologies, 
characterized by the existence of phononic edge and corner modes and by Wilson loops with winding numbers one and two. 
Fitting the dynamical matrix to the DFT phonon bands shows that graphene is not very far from a topologically nontrivial phonon phase.
\end{abstract}
\maketitle

\section{ Introduction} The honeycomb lattice  has played an important role in our understanding of topological states of matter. The  Haldane~\cite{PhysRevLett.61.2015} and Kane-Mele~\cite{PhysRevLett.95.226801,PhysRevLett.95.146802} models have clarified the effects of time-reversal symmetry, or its absence, on the topology of electron hamiltonians. For phonons, rotating honeycomb lattices~\cite{Wang_2015,Kariyado:2015kq,PhysRevB.96.064106}  where the Coriolis force plays the role of a  magnetic field and gyroscopic phononic crystals~\cite{PhysRevLett.115.104302} have been used to study the existence of nontrivial phonon topology in the absence of time-reversal symmetry. More recently, topological phonons have been reported on the valley-mixing Kekul\'e deformation of the honeycomb lattice~\cite{PhysRevLett.119.255901}.

On the other hand, new insights about band theory based on symmetry indicators~\cite{Po:2017qf,PhysRevX.8.031070}, band combinatorics~\cite{PhysRevX.7.041069} and Topological Quantum Chemistry (TQC)~\cite{Bradlyn:2017qf}  have been recently  used to classify all the nontrivial electron band topologies compatible with a given crystal structure. In particular, the methods of TQC have been used to uncover thousands of new materials with topological electron bands~\cite{2018arXiv180710271V}. But these new insights have not, to the best of our knowledge, been applied to the search for nontrivial  topology in  phonon spectra~\cite{2015Sci...349...47S,2016PNAS..113E4767S,Huber:2016fk,Kane:2013uq,PhysRevLett.117.068001,PhysRevB.86.035141,PhysRevLett.120.016401,PhysRevB.97.054305,PhysRevB.97.155422,PhysRevMaterials.2.114204}. In this paper we adapt the methods of TCQ to the analysis of phonon bands and, as a proof of principle, use them to unveil the existence of four previously unknown topological  phases on the time-reversal symmetric planar honeycomb. These phases are characterized by the existence of edge and corner phonon modes and nontrivial windings in the Wilson line spectra. 

\begin{figure}[h]
\begin{center}
\vspace*{-.4cm}
\includegraphics[angle=0,width=1.\linewidth]{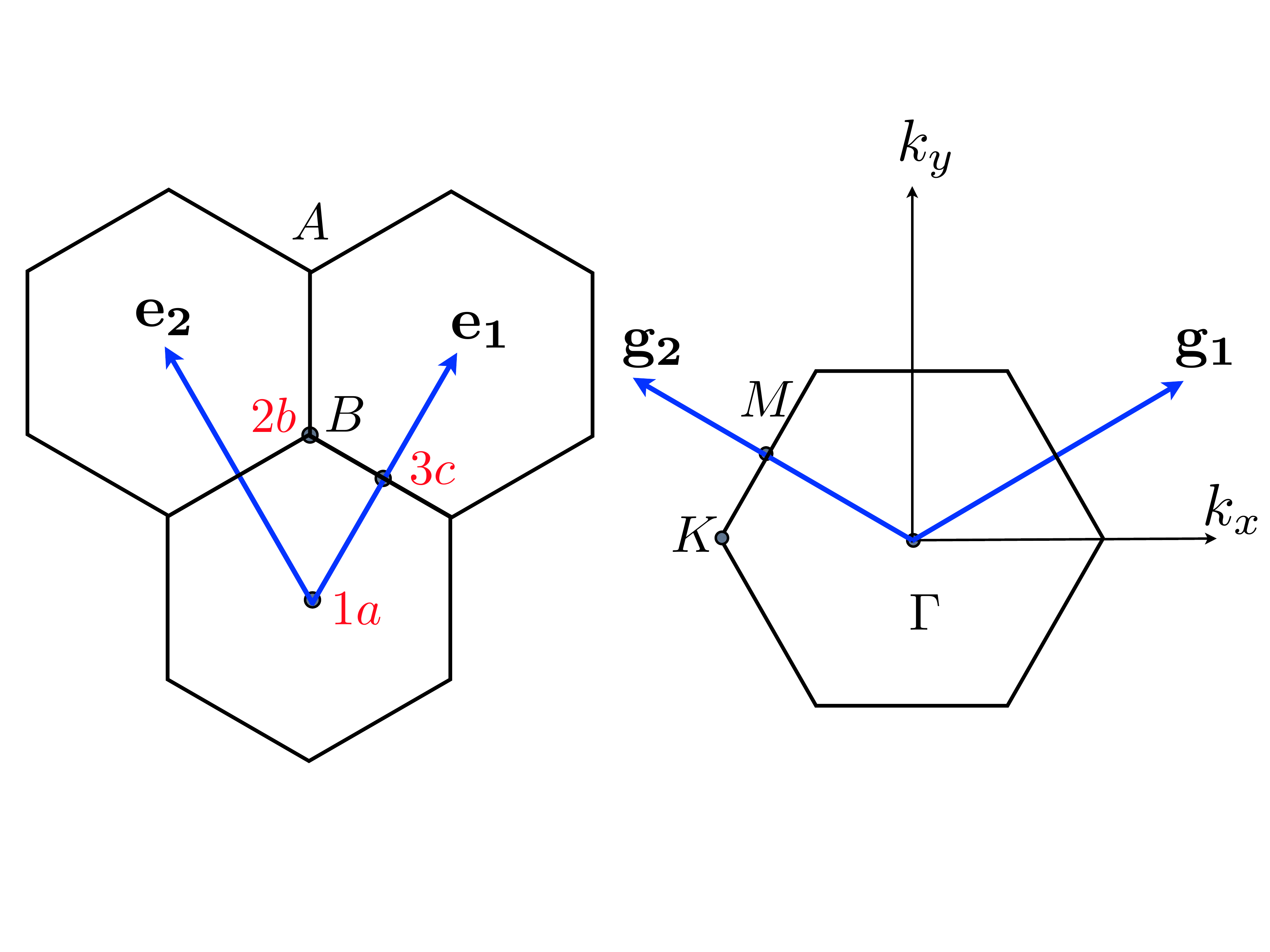}
\end{center}
\vspace*{-1.8cm}
\caption{The honeycomb lattice with basis vectors, maximal Wyckoff positions and high symmetry points in the BZ. The site symmetry groups are $C_{6v}$(1a), $C_{3v}$(2b) and $C_{2v}$(3c).}
\label{fig:f1}
\end{figure} 

 One key concept in the TQC framework is that of (elementary) band representation~\cite{PhysRevLett.45.1025,PhysRevB.23.2824,PhysRevB.97.035139}. Roughly speaking, a  band representation describes how  the set of atomic orbitals  in a crystal transform under the symmetry operations of the space group of the crystal. As the number of atoms  in an ideal crystal is infinite, a band representation is an infinite dimensional representation of the space group. The Bloch-wave combinations \hbox{$\sum_\mathbf{R} \exp(i\mathbf{k}\cdot\mathbf{R})\phi_i(\mathbf{r}-\mathbf{R})$} of  atomic orbitals~\cite{foot1} transform according to a representation of the little group $G_{\mathbf{k}}$~\cite{Bradley}. Thus a band representation induces little group representations at all the points in the Brillouin zone~(BZ).

A band representation that can not be written as the sum of two band representations is an elementary band representation (EBR). With some exceptions that are well understood and tabulated~\cite{MICHEL2001377,PhysRevB.97.035139}, an EBR is generated  from a set of orbitals that transform under an irreducible representation (irrep) $D$ of the local site symmetry group $G_w$ of a maximal symmetry Wyckoff position (WP)~$w$. The corresponding EBR will be denoted by~\hbox{$D@w$}.

Any band representation is either elementary or can be written as a sum of EBRs. If a subset of bands is separated by a gap from the other bands and does not transform as a band representation, then the subset does not have an atomic limit and is topologically nontrivial~\cite{Bradlyn:2017qf}. This gives an efficient method to identify topologically nontrivial subsets of bands. In practical terms, one considers the irreps describing the transformation of the subset of bands at the high symmetry points in the Brillouin zone. If the collection of little group irreps can not be obtained from a sum of EBRs, then the subset of bands necessarily has nontrivial topology. 

This method is applied in Section~\ref{sec:bandrep} to the phonon bands on the honeycomb, and a dynamical matrix that incorporates all the symmetry-allowed couplings including third nearest neighbors is presented in Section~\ref{sec:dynmat}. The phase diagram for the dynamical matrix is explored in Section~\ref{sec:phases}, where it is shown that, depending on the values of the coupling constant, the model can describe  four distinct fragile topological phonon phases. The fact that the dynamical matrix includes all the couplings compatible with the system symmetries allows for very precise fits to real materials. As an example, we place graphene on the phase diagram and show that it is not very far from one of the four fragile topological phases. The nontrivial topological phases are characterized by their Wilson loop winding  in Section~\ref{sec:wlwsym}, where it is shown that zero winding number is forbidden by symmetry for any isolated two-band branch. Edge and corner modes are analyzed in Section~\ref{edgecorner}, and the conclusions and discussion can be found in Section~\ref{discussion}.

\section{Band representations for phonons}\label{sec:bandrep} 
The concept of band representation is easily extended to the atomic vibrations of a crystal~\cite{PhysRevLett.74.4871,PhysRevLett.74.3824}. Instead of atomic orbitals, one considers the set of  vectors giving the  displacements from  equilibrium of all the atoms in the crystal and their transformation properties. This defines the ``mechanical'' band representation. Intuitively, the mechanical band representation can be understood as an electron band representation with (spinless) $p$-orbitals, since \hbox{$p$-orbitals}, like atomic displacements, transform according to the ``vector representation"~\cite{Bradley} $V$. Thus the mechanical band representation $M$  is induced from the vector representation at each occupied Wyckoff position $w$ in the crystal i.e., $M=\sum_w V@w$.

An important difference between electron and phonon band representations concerns the number of degrees of freedom and their location. Electronic orbitals are  functions   defined throughout space, and we can consider many orbitals in an atom, while an atomic displacement is described just by a single vector at the atom location. As a consequence, the number of phonon bands is always equal to three times the number of atoms in the primitive unit cell.  In this regard,  the  electromagnetic fields considered in the recent application of TQC to photonic crystals~\cite{2019arXiv190302562B} are closer to the electron wave-functions than to the crystal vibrations studied here. Another property of phonon bands without  parallel in  electrons is the existence of three acoustic bands, that must satisfy the constraint $\lim_{\mathbf{k}\to 0} \omega(\mathbf{k})=0$. The acoustic modes at $\mathbf{k}=0$ represent uniform translations $\mathbf{t}$ of all the atoms in the crystal and   transform according to the vector representation~$V$.

\subsection{Mechanical band representation for the planar honeycomb} The layer group for the planar honeycomb is LG 80 ($p6/mmm$), which corresponds to the space group SG 191 ($P6/mmm$). In-plane $(x,y)$ and off-plane $(z)$ vibrations of the lattice are respectively even and odd under reflections on the horizontal mirror plane, and   decouple in the harmonic approximation. Thus, for practical purposes, it is sufficient and simpler  to consider the subgroup SG 183 ($P6mm$), while treating in-plane and off-plane vibrations separately. Note, however, that the horizontal mirror plane symmetry is \textit{not} broken. This means, in particular, that we are considering a suspended lattice as opposed to a lattice on a substratum.

The atoms in the honeycomb lattice are at  the Wyckoff position $2b$, with site symmetry group isomorphic to the point group $3m$ ($C_{3v})$.
The vector representation for  $3m$  is reducible, $V=A_1(z)+E(x,y)$~\cite{Bradley}. This implies that the mechanical band representation can be written as the sum of two EBRs induced respectively from $A_1$ and $E$ at  $2b$, i.e., $M=A_1@{2b}+E@{2b}$, that  describe the transformation properties of the two off-plane  and four in-plane bands.

The application \textit{BandRep} at the Bilbao Christallographic Server (BCS)~ \cite{2006ZK....221...15A,2006AcCrA..62..115A,BulgChemCom43.183}   gives the irreps induced by any EBR at the high symmetry points in the BZ, and also tells whether the EBR is decomposable~\cite{PhysRevB.97.035138,PhysRevE.96.023310}.  An EBR is decomposable if the corresponding band can be split into two sets separated by a gap. It is known that a split EBR always gives rise to a topologically nontrivial phase~\cite{Bradlyn:2017qf,2018arXiv180409719B}. According to the BCS, $M=A_1@{2b}$ is indecomposable. In that sense the two off-plane phonon bands are nontopological  and will not be  considered further in this letter. Note, however, that they host a massless Dirac point at the $K$-point. On the other hand, $E@{2b}$ is decomposable and, as a consequence, the four in-plane bands can  split into two disconnected sets. Table~\ref{table:t1} gives, for all the subsets consistent with the compatibility relations,  the irreps at the three high symmetry points in the BZ. The acoustic branches are identified by noting that the vector representation at the $\Gamma$ point is given by $V=\Gamma_1(z)+\Gamma_6(x,y)$ and therefore $\Gamma_6$ is  associated with uniform in-plane translations.   
\begin{table}[h]
%\vspace*{.4cm}
\begin{tabular}{|c || l | | l |  }
\hline
Phase & Acoustic branch& Optical branch\\
\hline
\hline
Ia &  $\Gamma_6;K_1+K_2;M_1+M_2$ & $\Gamma_5;K_3;M_3+M_4$ \\
\hline
Ib &  $\Gamma_6;K_3;M_1+M_2$ & $\Gamma_5;K_1+K_2;M_3+M_4$ \\
\hline
IIa &  $\Gamma_6;K_1+K_2;M_3+M_4$ & $\Gamma_5;K_3;M_1+M_2$ \\
\hline
IIb &  $\Gamma_6;K_3;M_3+M_4$ & $\Gamma_5;K_1+K_2;M_1+M_2$ \\
\hline \end{tabular}
\caption{Irreps at the three high symmetry points in the BZ. All irreps are 1-dimensional, except for $\Gamma_5$, $\Gamma_6$ and $K_3$, which are 2-dimensional.
The little point groups at the $\Gamma$, $K$ and $M$ points are given by $6mm(C_{6v})$, $3m(C_{3v})$ and $C_{2v}(2m)$ respectively  }
\label{table:t1}
%\vspace*{-.3cm}
\end{table}

The next step is to determine which of the branches in Table~\ref{table:t1} can transform as band representations. To this end, we try to obtain  sums of EBRs that induce the same little group irreps at the high symmetry points in the BZ. It turns out that this is possible only for the optical branch of phase  IIa and the acoustic branch of IIb. As shown in Table~\ref{table:t2}, in order to obtain the irreps of the remaining branches  we must subtract some EBRs. This means that only the optical branch of phase  IIa and the acoustic branch of IIb may transform as band representations, whereas the remaining six branches must be topologically nontrivial. This is consistent with the general result that says that, when a set of bands transforming as an EBR splits, at least one of the resulting subsets of bands must have nontrivial topology~\cite{PhysRevLett.121.126402,PhysRevLett.120.266401,PhysRevB.99.045140}. The negative integral coefficients in Table~\ref{table:t2} are indicators of fragile topology, where the subset of bands may be trivialized by the addition of  trivial bands~\cite{PhysRevLett.121.126402}.  %\begin{widetext}{\ }
\begin{table}[h]
\begin{tabular}{|c ||c || c |  }
\hline
Phase & Acoustic branch& Optical branch\\
\hline
\hline
Ia &  $A_1@{1a}-A_1@{2b}+B_2@{3c}$ & $A_1@{3c}-A_1@{1a}$ \\
\hline
Ib &  $B_1@{3c}-B_2@{1a}$ & $B_1@{1a}+E@{2b}-B_2@{3c}$ \\
\hline
IIa & $E@{2b}-E_2@{1a}$ & $E_2@{1a}$ \\
\hline
IIb &  $E_1@{1a}$ & $E@{2b}-E_1@{1a}$ \\
\hline \end{tabular}
\caption{Combinations of EBRs that reproduce the irreps at the high symmetry points for the eight branches in Table~\ref{table:t1}. Note that these combinations are in general non-unique.}
\label{table:t2}
\end{table}
%\end{widetext}

The different phases in Table~\ref{table:t1} are associated with the ordering of the phonon frequencies at points $K$ and $M$. 
For example, in order to obtain phase Ia, $\omega(K_1)$ and 
$\omega(K_2)$ must be lower than $\omega(K_3)$, and $\omega(M_1)$ and $\omega(M_2)$ must also be lower than $\omega(M_3)$ and $\omega(M_4)$. This must be so in order to avoid band crossings between the two subsets, as such band crossings would cause the  subsets to reconnect~\cite{PhysRevB.97.035138}. For that reason, whenever $\omega(K_3)$ is between $\omega(K_1)$ and $\omega(K_2)$, the four bands are interconnected and we are in a nontopological phase. The frequencies at  different points in the BZ are obtained by diagonalizing the dynamical matrix~\cite{RevModPhys.40.1}.

\section{Dynamical matrix for in-plane modes}\label{sec:dynmat}  
The harmonic potential energy of the ions can be written as 
\beq
U=\frac{1}{2}\sum_{i,j}  \mathbf{r}_i^t   \mathrm{U}_{ij}\mathbf{r}_j,
\eeq
where $i,j$ run over all the atoms in the lattice, \hbox{$\mathbf{r}_i=(x_i,y_i)$} is the displacement of atom $i$ from  equilibrium and $U_{ij}=U_{ji}^t$ is the matrix of force constants between atoms $i$ and $j$. Assuming that the atoms $i,j$ are nth-nearest neighbors, the $2\times 2$ matrix $ \mathrm{U}_{ij}$ is  parametrized by four real coefficients $(a_n,b_n,c_n,d_n)$, often with a single set of coefficients for all nth-neighbor pairs.  As explained in Appendix~\ref{pardyn},   some of these coefficients may be forced to vanish by the symmetries of the lattice. 

The  basis vectors  for the Bravais lattice are given by
\beq
\mathbf{e}_1 =  \frac{1}{2}\hat{\mathbf{x}}+\frac{\sqrt{3}}{2}\hat{\mathbf{y}}\;\;,\;\;
\mathbf{e}_2 = -\frac{1}{2} \hat{\mathbf{x}} +\frac{\sqrt{3}}{2}\hat{\mathbf{y}},
\eeq
and are shown in Fig.~1 together with
their reciprocal lattice vectors $\mathbf{g}_i$, which satisfy $\mathbf{g}_i\cdot \mathbf{e}_j = 2\pi \delta_{ij}$.
Sites on the $A(B)$ sublattice are at positions $\mathbf{R} + \boldsymbol{\delta}_{A(B)}$, where $\mathbf{R}$ denotes a lattice translation and 
\beq
\boldsymbol{\delta}_A = \frac{2}{3}\mathbf{e}_1 + \frac{2}{3}\mathbf{e}_2\;\;,\;\;
\boldsymbol{\delta}_B = \frac{1}{3}\mathbf{e}_1 + \frac{1}{3}\mathbf{e}_2.
\eeq
Then each atom is specified by a composite label \hbox{$i\equiv(a, \mathbf R)$} that gives the unit cell $\mathbf R$ and the sublattice $a=A,B$.
In what follows it will be convenient to make a change of coordinates 
from linear to circular polarizations
\beq
\xi=\frac{1}{\sqrt{2}}(x+iy)\;,\; \bar\xi=\frac{1}{\sqrt{2}}(x-iy).
\eeq 

As shown in Appendix~\ref{pardyn}, in these coordinates, the matrix $U_{ij}$ can be written as
\beq
U_{ij}=a_n\openone_s+b_n s_x+c_n s_y+ id_ns_z,
\eeq
where $s_i$ are Pauli matrices for phonon `spin', with the upper (lower) components giving the amplitude of the right (left) circular polarization.
The dynamical matrix is obtained as  the Fourier transform of the matrix of force constants
\beq\label{ft}
D_{ab}(\mathbf{k})=\sum_{\mathbf{R}}U_{a,\mathbf{0};b,\mathbf{R}}\,e^{i\mathbf{k}\cdot\mathbf{R}}.
\eeq
Note that the vector indices $(x,y)$ or $(\xi, \bar\xi)$ are implicit.

The dynamical    matrix acts on 4-component vectors 
\beq\label{basis4} 
v(\mathbf{k})^t=(\xi_A(\mathbf{k}),\bar\xi_A(\mathbf{k}),\xi_B(\mathbf{k}),\bar\xi_B(\mathbf{k}))
\eeq
and can be expressed in terms of Kronecker products of Pauli matrices $\sigma_i$ for the two sublattices, and $s_i$ for phonon `spin'.
The result, up to third nearest neighbor interactions is given by
\vskip-.9cm
\begin{widetext}{\ }
\begin{align}\label{dynmat}
D(\mathbf{k})=&-3(a_1+2a_2+a_3)\openone_4+a_1\left[ \left(1+2\cos\x\cos  \y\right)\sigma_x\otimes \openone_s  -2\left(\cos  \x\sin  \y\right)\sigma_y\otimes \openone_s\right]\non\\
&+b_1\left[ \left( 1-\cos\x\cos \y\right) \sigma_x\otimes s_x+\left( \cos\x\sin \y\right) \sigma_y\otimes s_x\right. \non\\
&\left.-\sqrt{3}\left( \sin\x\cos\y\right) \sigma_y\otimes s_y -\sqrt{3}\left( \sin\x\sin\y\right) \sigma_x\otimes s_y
\right]\non
\end{align}
\begin{align}
&+a_2\left(  2\cos\xx+ 4\cos\x\cos\y \right)\openone_4+d_2\left( 4\sin \x\cos \y -2\sin\xx \right)\sigma_z\otimes s_z\non\\
&+b_2\left[ 2\left( \cos\xx-\cos\x\cos\y\right)\openone_\sigma \otimes s_x +2\sqrt{3}\left(\sin\x\sin\y\right)\openone_\sigma \otimes s_y   \right]\non\\[.2cm]
&+a_3\left[\left( 2 \cos\xx +\cos\yy\right)\sigma_x\otimes \openone_s  -(\sin\yy)\,\sigma_y\otimes \openone_s\right]\non\\[.3cm]
&+b_3\left[\left( -\cos\xx+\cos\yy \right)\sigma_x\otimes s_x-(\sin\yy)\,\sigma_y\otimes s_x +\sqrt{3}(\sin\xx)\,\sigma_y\otimes s_y \right].
\end{align}
\end{widetext}
Note that all the coefficients in this expression are real.  A detailed analysis of the symmetries of the dynamical matrix can be found in Appendix~\ref{sec:symdyn}.

It is interesting to note that, if we neglect third nearest neighbor interactions, the dynamical matrix~(\ref{dynmat}) is closely related to the Kane-Mele hamiltonian~\cite{PhysRevLett.95.146802} for spinful $p_z$ orbitals in graphene, with the two phonon circular polarizations playing, to some extent, the role of electron spin. In fact, comparing~(\ref{dynmat}) with the Kane-Mele hamiltonian shows that the terms proportional to $a_1$ have the structure of the Dirac hamiltonian, while the term with $d_2$ coincides with the Haldane spin-orbit coupling  in the Kane-Mele model. On the other hand, the functions of ($k_x,k_y$) in the terms proportional to $b_1$ are identical to the ones appearing in the Rashba spin-orbit coupling  in the Kane-Mele model, although the matrix structures are different. Actually, writing the couplings in real space shows that the Rashba term in the Kane-Mele model differs from  the $b_1$-term just by a factor of the imaginary unit $i$. That both terms manage to be time-reversal invariant reflects the fact that electron spin-$1/2$ and phonon spin-$1$ transform differently under time reversal.  

Another consequence of the different transformation properties of electron and phonon spin, this time under space symmetries,  is that the Rashba coupling breaks reflection symmetry by the horizontal mirror plane, while the $b_1$-term does not. Finally, one last manifestation of the differences between spin-$1/2$ and spin-$1$ systems is the absence of Kramers degeneracy in the phonon spectrum, where we find four different frequencies at the time-reversal symmetric $M$-point, instead of the two doubly degenerate energies of the electron spectrum.

As mentioned above, in-plane displacements transform according to the vector representation, just  like  $(p_x,p_y)$ orbitals do. As a consequence, the dynamical matrix for in-plane modes has to be closely related to the hamiltonian for spinless $(p_x,p_y)$-orbitals in graphene. This is  actually  the case, as the model in Ref.~\onlinecite{PhysRevLett.120.266401} coincides with the dynamical matrix~(\ref{dynmat}) if we take \hbox{$(a_1,b_1)\sim (t_\sigma\pm t_\pi)$},  $d_2\sim x$, where $(t_\sigma,t_\pi, x)$ are the couplings in Ref.~\onlinecite{PhysRevLett.120.266401}, and set all the other couplings and the on-site term (proportional to $\openone_4$) in~(\ref{dynmat}) to zero. Note, however, that in that limit phases IIa and IIb, to be discussed in the next section, are out of reach and therefore were not mentioned in Ref.~\onlinecite{PhysRevLett.120.266401}.

\section{Phase diagram for the planar honeycomb}
\label{sec:phases}
The dynamical  matrix can be diagonalized analytically at the three high symmetry points in the BZ, and  irreps can be assigned to the corresponding eigenmodes. The result may be written as
\beq
\omega^2(\mathbf{k},D_a)=\omega^2(\mathbf{k})+\delta \omega^2(\mathbf{k},D_a),
\eeq
where $D_a$ denotes an irrep of the little group $G_{\mathbf{k}}$, $\omega^2(\mathbf{k})$ is common to all the bands at the $\mathbf{k}$-point and $\delta \omega^2(\mathbf{k},D_a)$ is specific to each irrep. We obtain
\begin{align}
\omega^2(\Gamma)&=0\non\\
\omega^2(K)&=-3(a_1+3a_2+a_3)\non\\
\omega^2(M)&=-3a_1-8a_2-3a_3,
\end{align}
while $\delta \omega^2(\mathbf{k},D_a)$ is given in Table~\ref{table:ts2}, where the last column shows the $C_{2z}$-eigenvalues of the normal modes at the $C_{2z}$-invariant points $\Gamma$ and $M$. %The $C_{2z}$-eigenvalues are obtained from the  matrices for the irreps of the little groups tabulated at the BCS.

\begin{table}[h]
\renewcommand{\arraystretch}{1.2}
\begin{tabular}{|c|c ||c | c | c |}
\hline
$G_{\mathbf{k}}$ & Irrep & d & $\delta \omega^2$ & $C_{2z}$\\
\hline
\hline
$C_{6v}$ & $\Gamma_5$ & $2$ & $-6(a_1+a_3)$ & $+1$ \\
 & $\Gamma_6$ & $2$ & $0$ & $-1$ \\
\hline
$C_{3v}$ &$K_1$ & 1 & $3\sqrt{3} d_2+3(b_1+b_3)$ & -\\
&$K_2$ & 1 & $3\sqrt{3} d_2-3(b_1+b_3)$ & -\\
&$K_3$ & 2 & $-3\sqrt{3} d_2$ & -\\
\hline
$C_{2v}$ &$M_1$ & 1 & $a_1-3a_3+2 b_1-4 b_2$ & $+1$ \\
&$M_2$ & 1 & $a_1-3a_3-2 b_1+4 b_2$ & $+1$ \\
&$M_3$ & 1 & $-a_1+3a_3-2 b_1-4 b_2$ & $-1$ \\
&$M_4$ & 1 & $-a_1+3a_3+2 b_1+4 b_2$ & $-1$ \\
\hline

\end{tabular}
\caption{$\delta \omega^2(\mathbf{k},D_a)$ for the little group irreps. $d$ is the dimension of the irrep.  The last column indicates the \hbox{$C_{2z}$-eigenvalues} of the normal modes.}
\label{table:ts2}
\end{table}
As mentioned above, in order to be in a nontrivial phase $\omega(K_3)$ should  be  higher or lower than $\omega(K_1)$ and $\omega(K_2)$. A look at the frequencies in Table~\ref{table:ts2} shows  that this implies 
\beq\label{eq1}
2\sqrt{3} |d_2| > |b_1+b_3|.
\eeq
A second necessary condition in order to have two disconnected sets of bands is that $\omega(M_1)$ and  $\omega(M_2)$ are both higher or lower than $\omega(M_3)$ and  $\omega(M_4)$. This is equivalent to 
\beq\label{ineq}
|2b_1+4b_2|+|2b_1-4b_2|<2|a_1-3a_3|.
\eeq
For $|b_1|>2|b_2|$ this reduces to $2|b_1|<|a_1-3a_3|$ while, for $|b_1|<2|b_2|$, Eq.~(\ref{ineq}) is equivalent to $4|b_2|<|a_1-3a_3|$. The two cases can be combined into
\beq\label{eq2}
|a_1-3a_3|>\mathrm{Max} \{ 2|b_1|,4|b_2| \}.
\eeq
\begin{table}[h]
\begin{tabular}{| c | c c |}
\hline
Phase &   Conditions& \\
\hline\hline
 Ia& $d_2<0$ & $a_1<3a_3$  \\
Ib & $d_2>0$ & $a_1<3a_3$  \\
 IIa & $d_2<0$ & $a_1>3a_3$  \\
IIb & $d_2>0$ & $a_1>3a_3$  \\
\hline
\end{tabular}
\caption{Topologically nontrivial phases.  }
\label{table:ts4}
\end{table}
 
 Assuming that Eqs.~(\ref{eq1},\ref{eq2}) are satisfied, Table~\ref{table:ts4} identifies the nontrivial topological phases.
If either condition in Eqs.~(\ref{eq1},\ref{eq2}) fails to be satisfied, we will be in a nontopological phase where all four bands are interconnected.
It is important to note that,  unlike the hamiltonian $H(\mathbf{k})$ for electron bands, the dynamical matrix has to satisfy several stability conditions to prevent the existence of imaginary phonon frequencies. For instance, imposing $\omega^2(\Gamma) >0$  requires $a_1+a_3<0$.

Note also  that phases Ia and Ib can be modeled by a dynamical matrix including only up to second nearest neighbors, whereas  third nearest neighbor interactions are required to obtain phases IIa and IIb. 
Specifically, a minimal model that describes qualitatively  the topologically nontrivial phases with WL  winding number one (Ia and Ib, see next section) can be obtained by keeping only the set of parameters $(a_1, b_1, d_2)$,  with $ |a_1|>2|b_1|$ and  $2\sqrt{3} |d_2| > |b_1|$, with   $a_1$  negative for stability. Phases IIa and IIb can be obtained by keeping also $a_3$, with $|a_1-3a_3|>2|b_1|$ and $3a_3<a_1$. Note that this last condition requires $3|a_3|>|a_1|$ which, as shown at the end of this section, would require increasing $a_3$  by a factor of twenty in the case of graphene. Thus having $3|a_3|>|a_1|$   is probably  unrealistic for a material crystal, but may be attainable for mechanical systems or metamaterials~\cite{Maldovan:2013fk}. 

\subsection{Locating graphene on the phase diagram }
\label{sec:graphfit}
\begin{figure}[h]
\begin{center}
\includegraphics[angle=0,width=.49\linewidth]{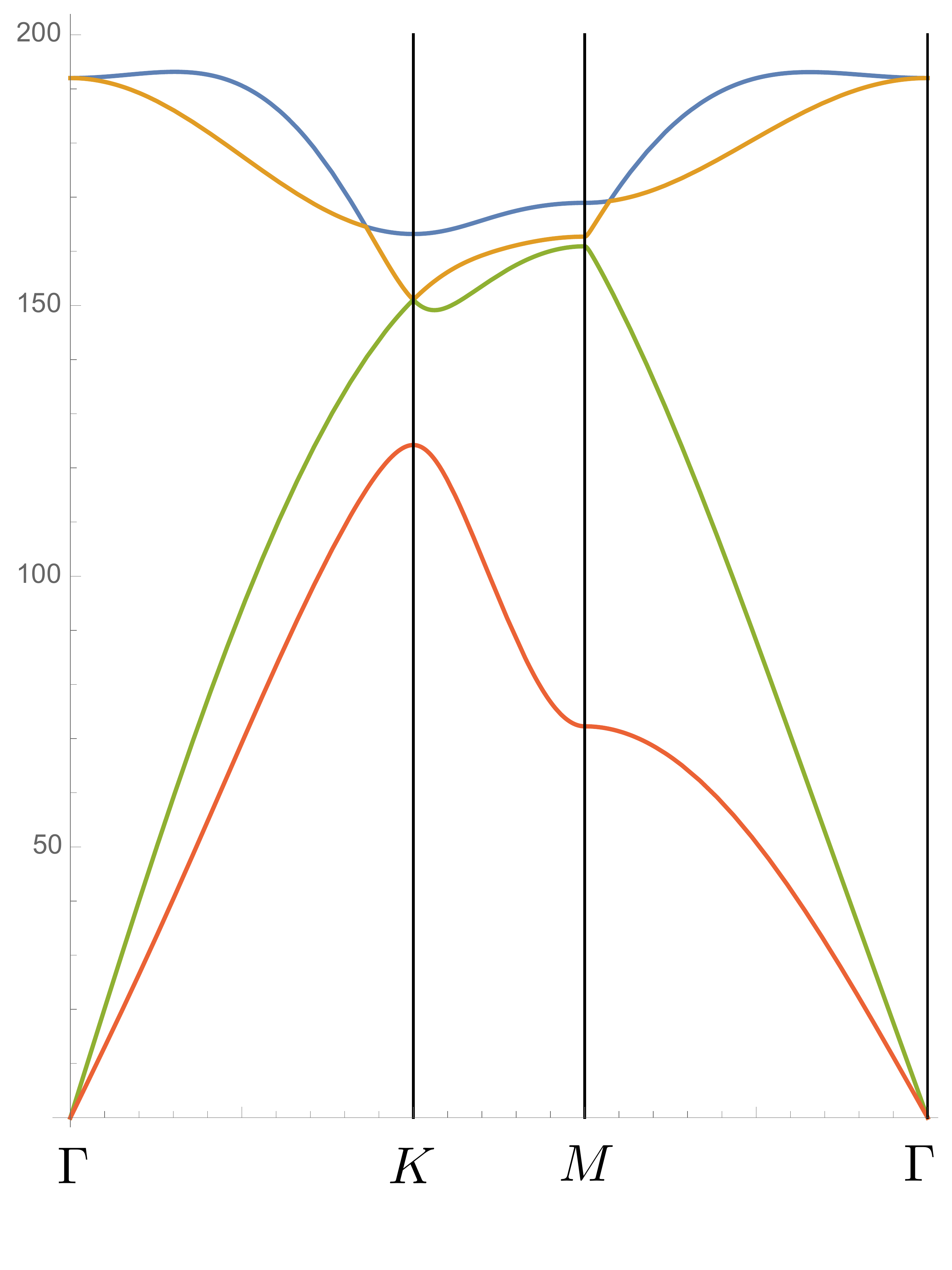}
\includegraphics[angle=0,width=.49\linewidth]{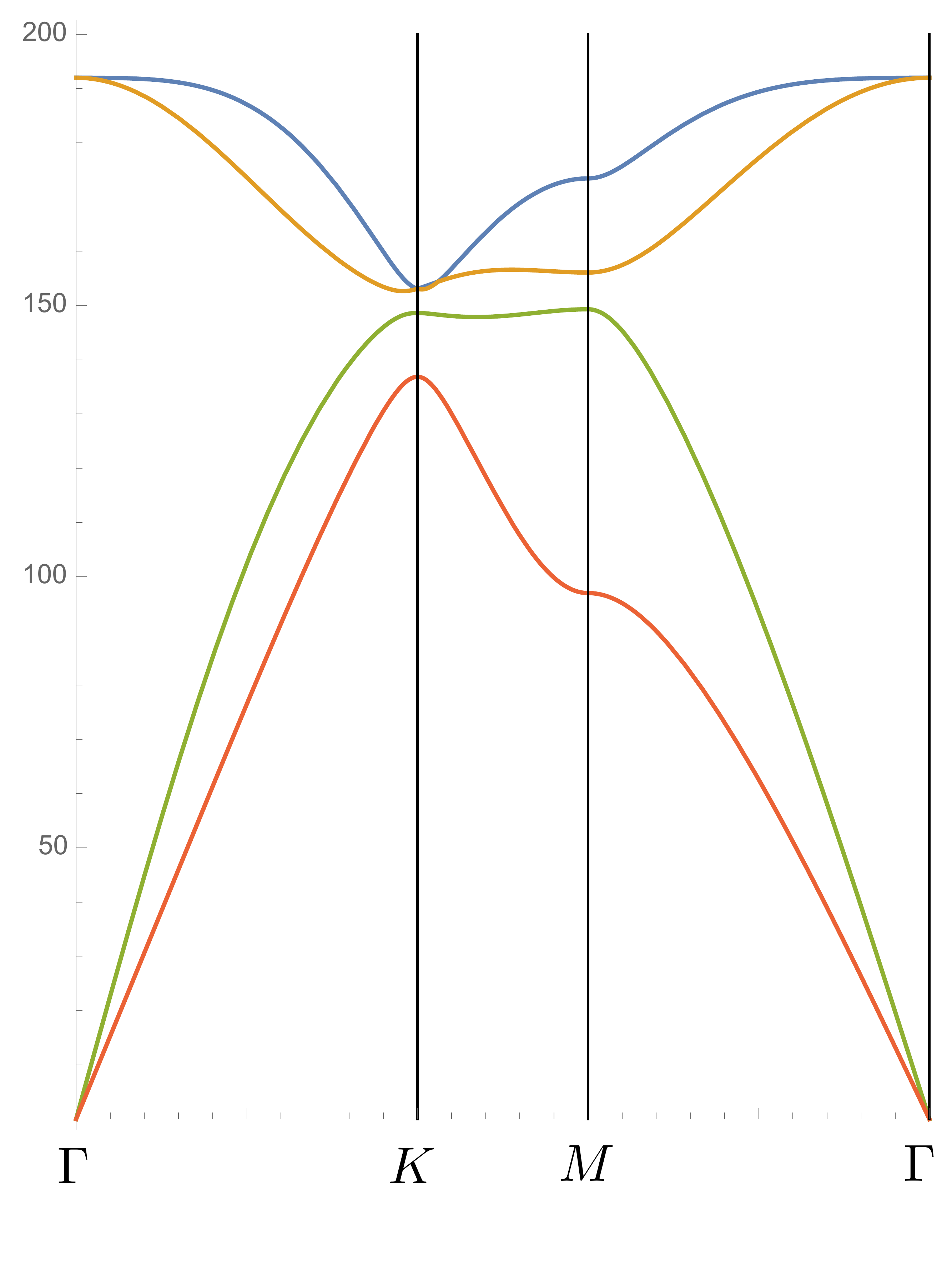}
\end{center}
\vspace*{-.8cm}
\caption{In-plane phonon spectrum for graphene, obtained by diagonalization of the third-nearest neighbor dynamical matrix. Left: Phonon bands for the values of the parameters that give the best fit to the DFT bands. Right: Phonon bands after increasing $d_2$ by 70\%, while decreasing $b_1$ and $b_3$ in the same proportion.}
\label{fig:f2}
\end{figure}

Here we will try to place graphene on the phase diagram for the in-plane phonon system of the planar honeycomb. To this end, we fit the DFT phonon spectrum reported in Ref.~\onlinecite{PhysRevLett.114.245502} to the 
 one obtained from the third-nearest neighbor dynamical matrix~(\ref{dynmat}).
\begin{table}[h]
\begin{tabular}{| c || c| c |}
\hline
 &   DFT& 3NN \\
\hline\hline
$ \omega(\Gamma_5)$ & $192 $ & $192 $  \\
\hline
$ \omega(K_1)$ & $122$ & $124$  \\
$ \omega(K_2)$ & $161$ & $163$  \\
$ \omega(K_3)$ & $149$ & $151$  \\
\hline
$ \omega(M_1)$ & $163$ & $161$  \\
$ \omega(M_2)$ & $171$ & $169$  \\
$ \omega(M_3)$ & $77$ & $72$  \\
$ \omega(M_4)$ & $165$ & $163$ \\
\hline 
\end{tabular}
\caption{Phonon energies in $meV$ at the high symmetry points. }
\label{table:ts5}
\end{table}
Table~\ref{table:ts5} compares the phonon frequencies at the high symmetry points computed by DFT in Ref.~\onlinecite{PhysRevLett.114.245502} with the ones obtained by diagonalizing the third-nearest neighbor dynamical matrix~(\ref{dynmat}) for the  following set of parameters, in units of $(meV)^2$ 
\beq\label{param}
\begin{array}{clccl}
a_1&=-6.03\cdot 10^3 & & b_2&=1.16\cdot 10^3\\
a_2&=-0.387\cdot 10^3 & & b_3&=1.12\cdot 10^3\\
a_3&=-0.114\cdot 10^3 & & d_2&=-0.171\cdot 10^3\\
b_1&=-2.99\cdot 10^3. & & &\\
\end{array}
\eeq
The parameters have been obtained by using the explicit formulae in Table~\ref{table:ts2} to fit the in-plane bands in Fig.~2 of Ref.~\onlinecite{PhysRevLett.114.245502} at the high symmetry points. As these are not labeled by irreps, some trial and error is necessary before the fit can be completed. The goal  is to get the best possible agreement at the high symmetry points, while maintaining the overall  qualitative features of the spectrum. This leads to the unique assignment of irreps at the high symmetry points given in Table~\ref{table:ts5}. As shown in Table~\ref{table:ts4}, the energies at the high symmetry points agree within a few percent, and the spectrum in Fig.~\ref{fig:f2} (left) closely resembles the one in  Ref.~\onlinecite{PhysRevLett.114.245502}.

All the in-plane bands in graphene are interconnected and the spectrum is nontopological. But now that we have parametrized the spectrum as in~(\ref{param}), we can try to see how far  graphene is  from becoming topological. Plugging~(\ref{param}) into~(\ref{eq1})  yields
\beq
2\sqrt{3} |d_2| \simeq 0.31|b_1+b_3|,
\eeq
while for~(\ref{eq2}) the result is
\beqa
|a_1-3a_3|&\simeq 0.95\cdot 2|b_1|\non\\
|a_1-3a_3|&\simeq 1.22\cdot 4|b_2|.
\eeqa 
Thus we see that, whereas  inequality~(\ref{eq2}) is on the verge of being satisfied, we are further  from satisfying~(\ref{eq1}), although   the couplings are at least the correct order of magnitude. 

How much do we need to vary the parameters to get into a topologically non-trivial phase? Assume that we increase the strength of $d_2$ in the same proportion that we decrease $b_1$ and $b_3$, i.e., we take $d_2\to(1+x)d_2$, together with $b_1\to(1-x)b_1$ and $b_3\to(1-x)b_3$. Then we see that the required inequalities are satisfied for \hbox{$x\gtrsim .52$}. In other words, graphene can enter a topologically nontrivial phonon phase if  some couplings are changed by about  $50\%$. As graphene can be stretched by up to $20\%$  and coupling constants are often exponentially sensitive to atomic distances, this is not a wholly unrealistic possibility. 
Fig.~\ref{fig:f2}(right) shows the in-plane phonon spectrum for $x=0.7$. As the frequency of $\omega(K_3)$ is greater than $\omega(K_1)$ and $\omega(K_2)$, we are in Phase Ia. Note that phases IIa and IIb are clearly  out of reach, as they would require increasing the strength of $a_3$ by a factor of twenty, from $|a_3|\simeq 0.11\cdot 10^3$ to more than $|a_1|/3=2 \cdot 10^3$.

\section{Wilson loop winding and symmetries} \label{sec:wlwsym}
\begin{figure}[h]
\begin{center}
\includegraphics[angle=0,width=1.\linewidth]{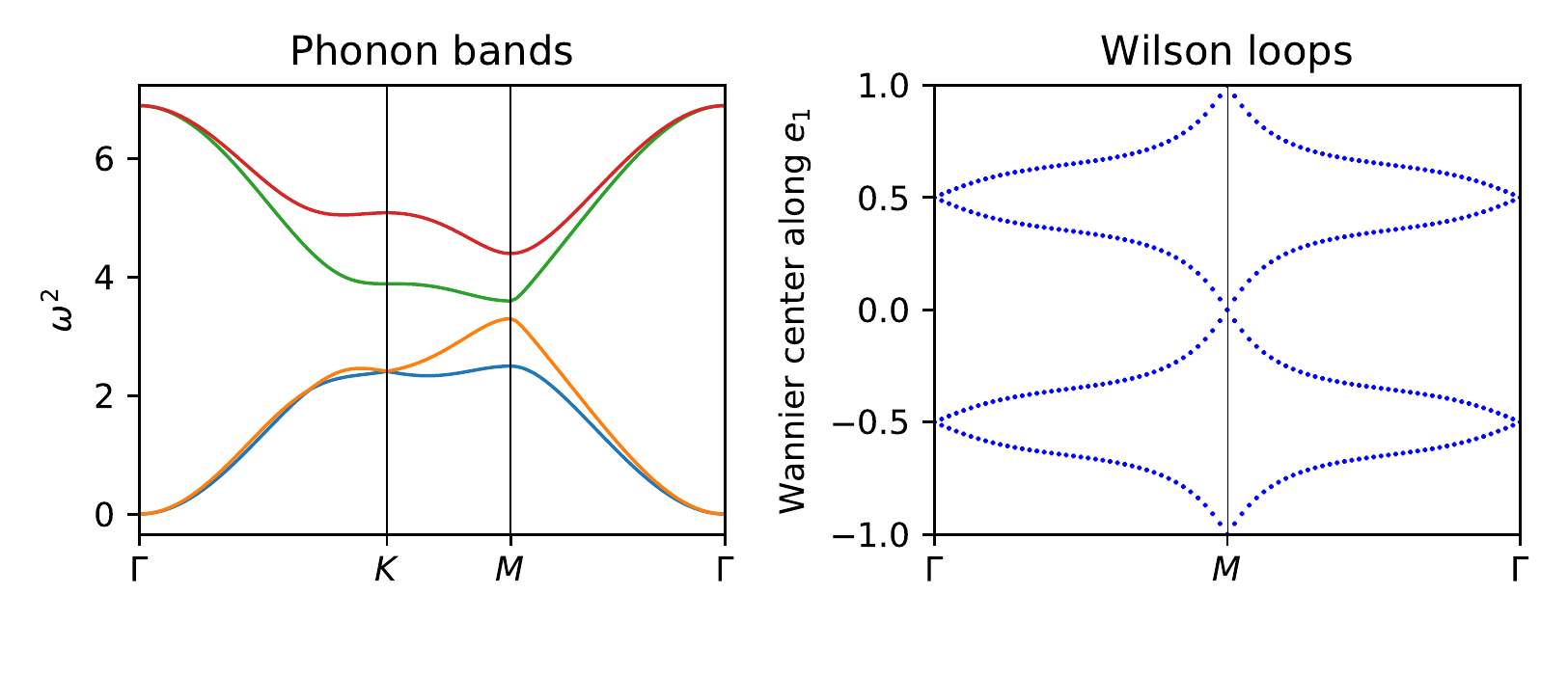}
\includegraphics[angle=0,width=1.\linewidth]{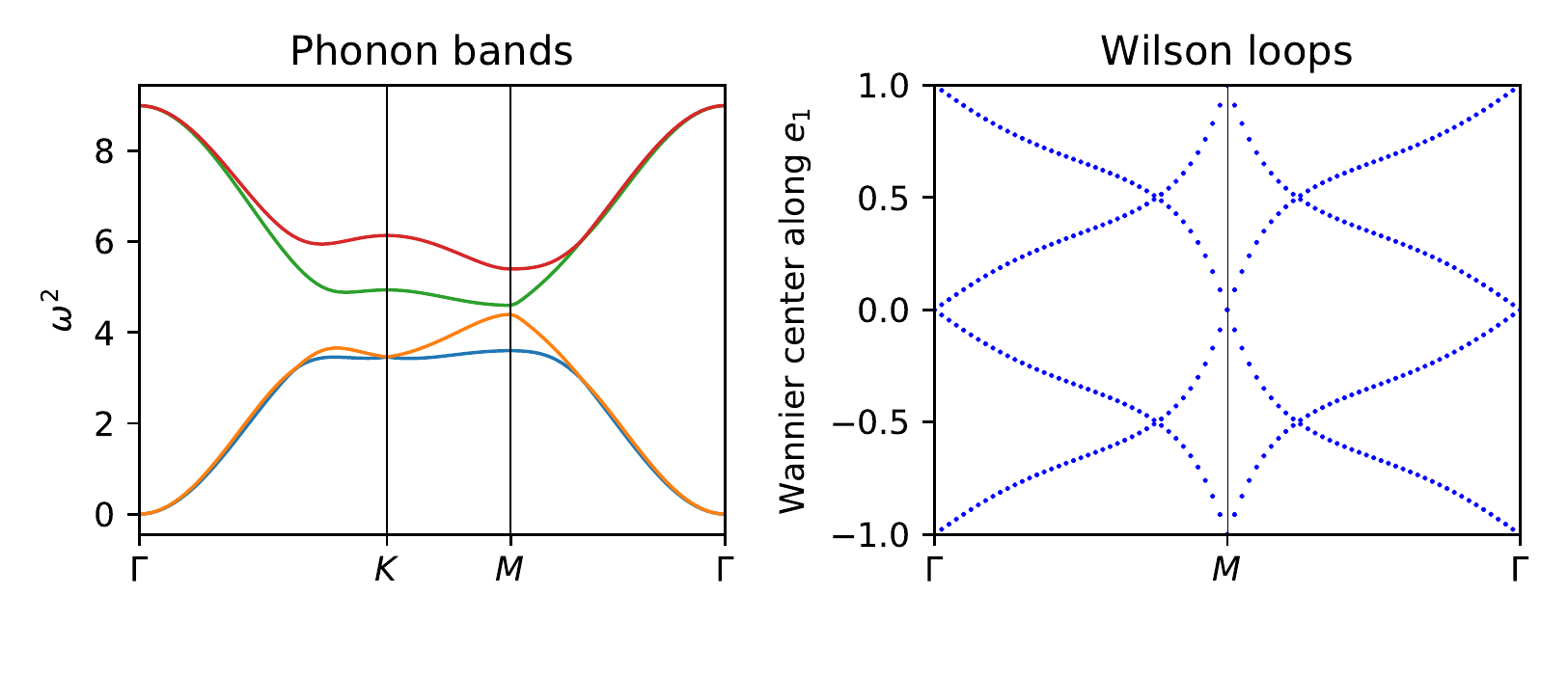}
\end{center}
\vspace*{-.8cm}
\caption{Phonon bands (left)  and Wannier centers for the acoustic bands (right). Top: $a_1=-1$, $a_3= -0.15$, $b_1=-0.2$, $d_2=0.2$, $a_2=b_2=b_3=0$ (Phase Ib). Bottom: $a_1=-1$, $a_3= -0.5$, $b_1=-0.2$, $d_2=0.2$, $a_2=b_2=b_3=0$ (Phase IIb) }
\label{fig:f3}
\end{figure}

Winding in the Wilson loop (WL) spectrum for a subset of isolated bands is a topological invariant. A winding that can not be eliminated by any perturbation that respects the symmetries of the system and does not close a gap guarantees  that  the subset of bands has nontrivial topology. We will consider a $\mathbf{g_1}$-directed Wilson loop~\cite{PhysRevLett.52.2111,PhysRevB.83.035108,PhysRevLett.48.359,PhysRevLett.62.2747,PhysRevX.6.021008,foots3} defined  by 
\beq
 W(k_2) \equiv  P e^{i\int_{0}^{2\pi} dk_1 A_1(k_1,k_2)},  
\label{eq:WL}
\eeq
where $P$ indicates that the integral is path-ordered and $A_1(\mathbf{k})_{ij} = i \langle u_i(\mathbf{k}) | \partial_{k_1} u_j (\mathbf{k})\rangle$ is the non-abelian Berry potential built from the normal modes $u_i(\mathbf{k})$ of a subset of isolated bands. We take $k_1$ along $\mathbf{g_1}$ (see Fig.~\ref{fig:f1}) and $k_2$ along $\mathbf{g_2}$. The eigenvalues of $W(k_2)$ are of the form $e^{2\pi i x_1(k_2)}$, where $x_1(k_2)$ is the position  of the center of a hybrid Wannier function~\cite{PhysRevB.26.4269,PhysRevB.56.12847} along  $\mathbf{e_1}$.  As  the base-point moves along the $k_2$-axis from $\Gamma$ to $M$ and back to $\Gamma$, the Wannier centers move along the 1-dimensional unit cell, as shown in Fig.~\ref{fig:f3}~\cite{foot7}.

According to Table~\ref{table:t2}, all the branches but the optical one of phase IIa and the acoustic one of IIb are necessarily topological, as they do not transform as  band representations. On the other hand, the irreps of the optical branch of phase IIa and the acoustic branch of IIb at the high symmetry points of the BZ are such that they \textit{might} transform according to  band representations and could be trivial. However, as shown in Fig.~\ref{fig:f3}, the WL of the acoustic branch of IIb has winding number two, which implies that the branch has nontrivial topology.
In fact, we find that the WLs of \textit{all} eight  branches in Table~\ref{table:t2} have non-zero winding numbers, equal to one for phases Ia and Ib, and two for IIa and IIb.
In other words, there are no trivial bands in the topological phases of this phonon system.

The windings of the WLs can be understood from the $C_{2z}$-eigenvalues of the normal modes at the two   \hbox{$C_{2z}$-invariant} points, $\Gamma$ and $M$. As shown in Table~\ref{table:ts2}, the eigenvalues are $+1$ for $\Gamma_5$, $M_1$ and $M_2$, and~$-1$ for $\Gamma_6$, $M_3$ and $M_4$. Comparing with Table~\ref{table:t1}, we see that the two bands in each branch of phases Ia and Ib have the same  $C_{2z}$-eigenvalues at the $\Gamma$-point, and opposite to those at the   $M$ point. The converse is true for phases IIa and IIb, where the $C_{2z}$-eigenvalues at the $\Gamma$ and $M$ points are equal. Having opposite $C_{2z}$-eigenvalues at $\Gamma$ and $M$ forces the Wilson bands to wind, while having the same eigenvalues is compatible with zero or, more generally,  even winding numbers~\cite{PhysRevB.89.155114}. The crossings  at generic $k_2$ and Wannier center $x_1=\pm1/2$ in Fig.~\ref{fig:f3} are protected by $C_{2z}\mathcal{T}$ invariance~\cite{2018arXiv180409719B,PhysRevB.99.045140,2018arXiv180710676S}, where $\mathcal{T}$ is the time-reversal operation. A more involved analysis including the role of the $C_{3z}$-rotation symmetry~\cite{2018arXiv180409719B,PhysRevB.99.045140,2018arXiv180710676S} shows that the allowed WL winding numbers are of the form $3n\pm 1$ ($n\in \mathbb{Z}$), see Appendix~\ref{sec:wlw} for details. Thus zero winding is excluded, which explains the absence of trivial bands in the topological phases of this system. 

We close this section by pointing out that our analysis of Wilson loop winding rests on the possibility of defining Wannier functions for phonons. In fact, the possibility  of defining localized basis for lattice vibrations was established long ago by W. Kohn~\cite{PhysRevB.7.2285} in a simple one-dimensional model, while a modern general treatment based on the extension of the concept of position operator to phonons can be found in Ref.~\onlinecite{PhysRevLett.96.216403}.

\section{Edge and corner modes}\label{edgecorner} 

\begin{figure}[h]
\begin{center}
\includegraphics[angle=0,width=.45\linewidth]{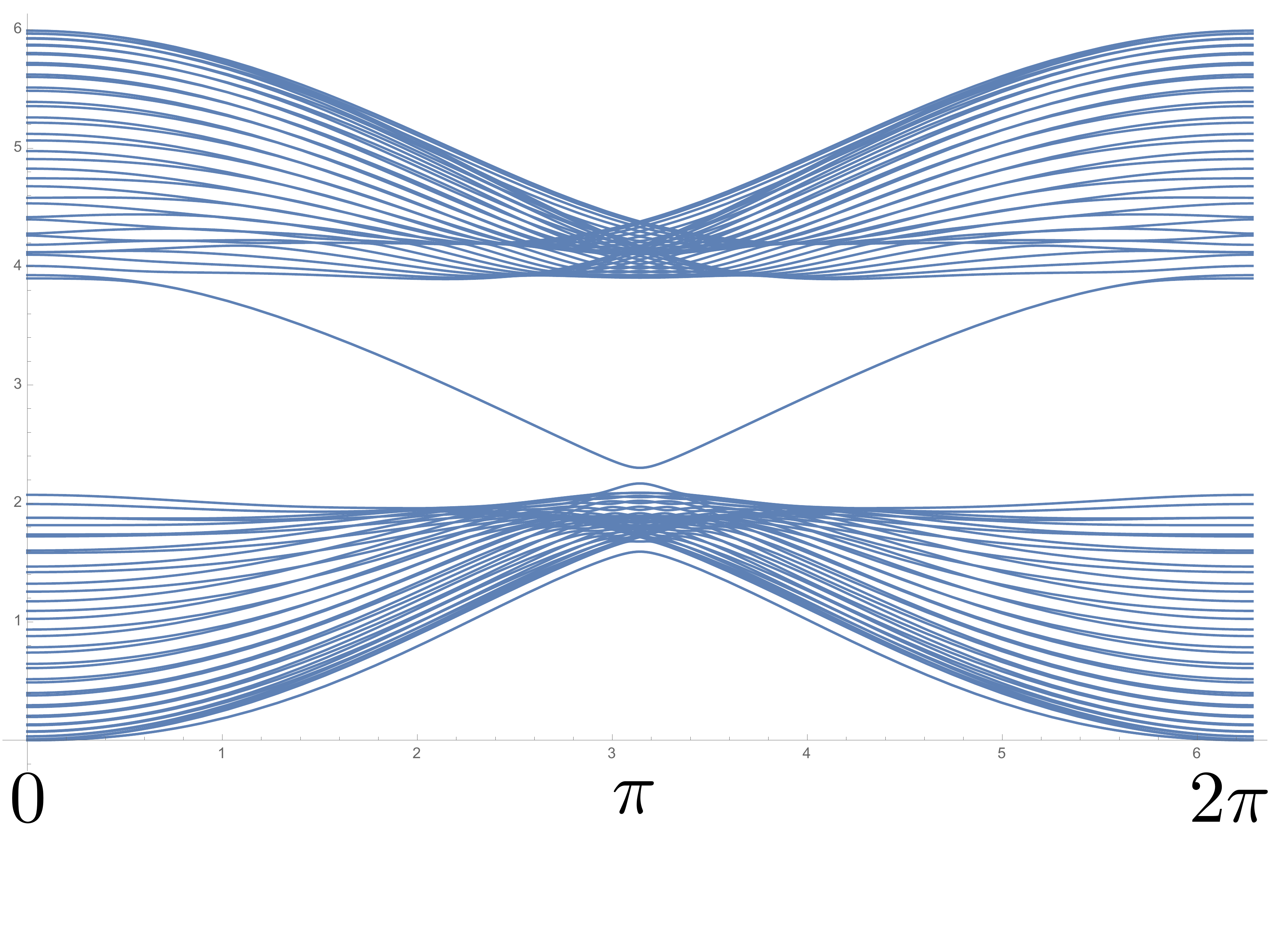}
\includegraphics[angle=0,width=.45\linewidth]{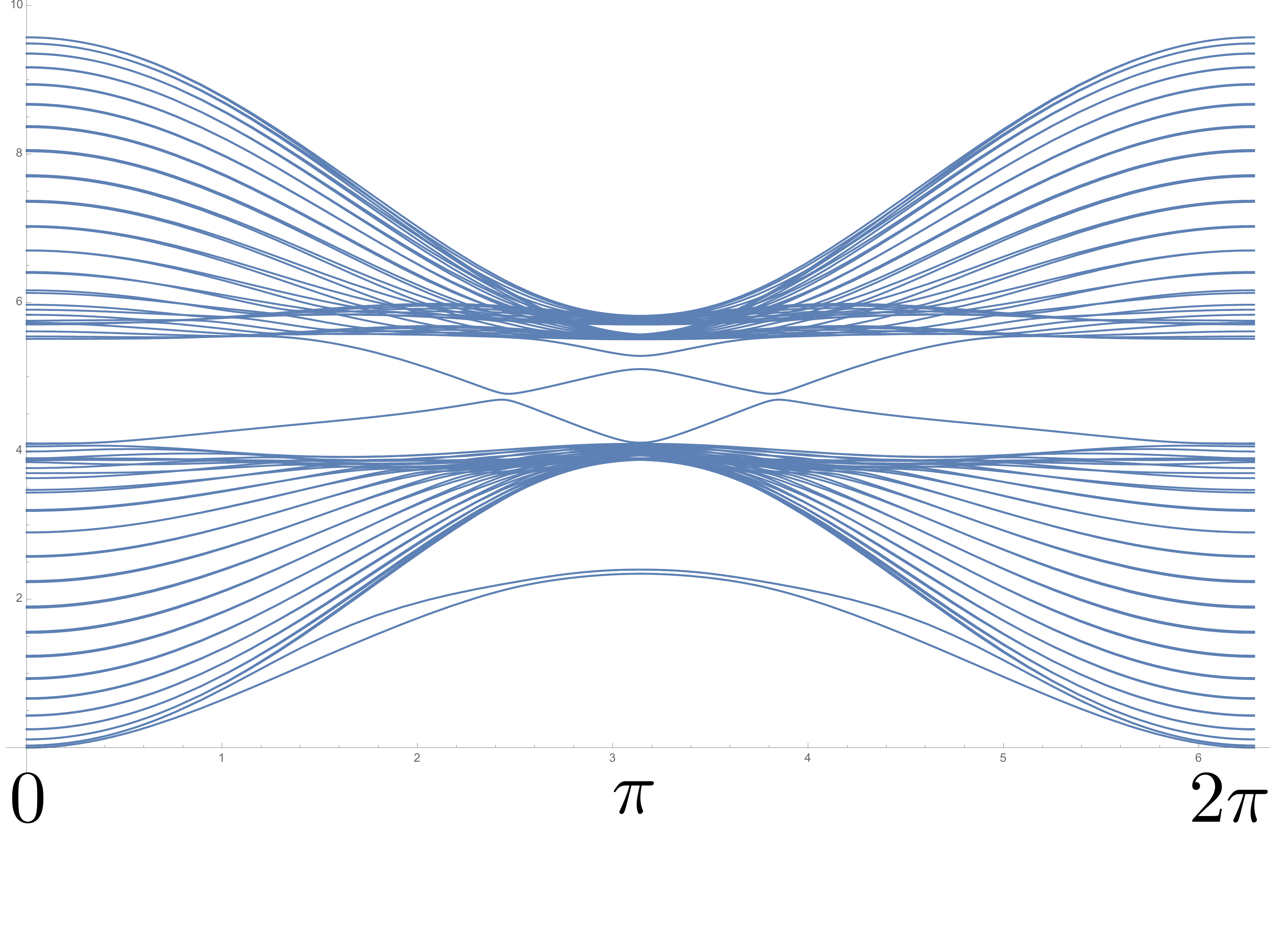}
\includegraphics[angle=0,width=.4\linewidth]{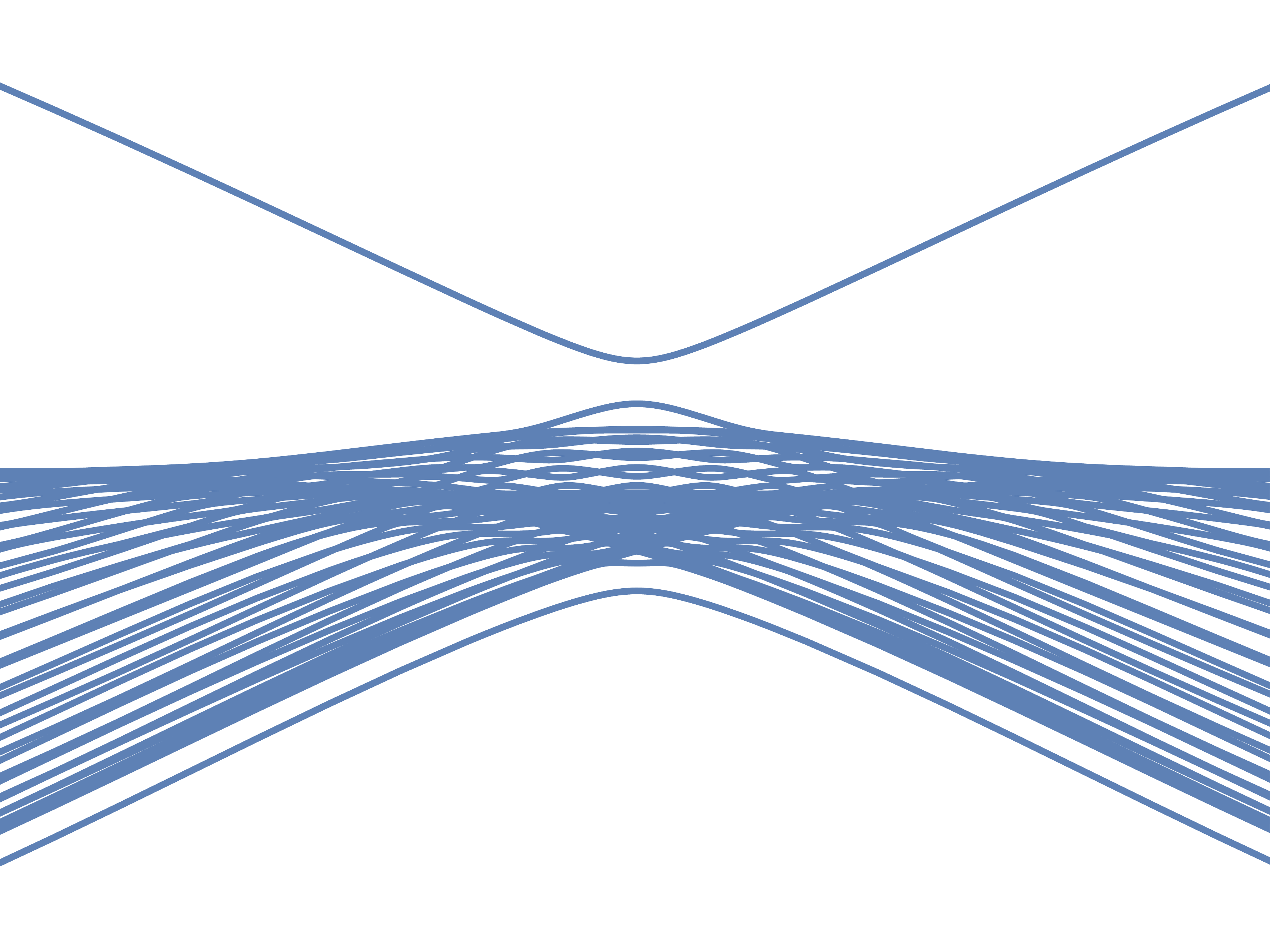}
\includegraphics[angle=0,width=.4\linewidth]{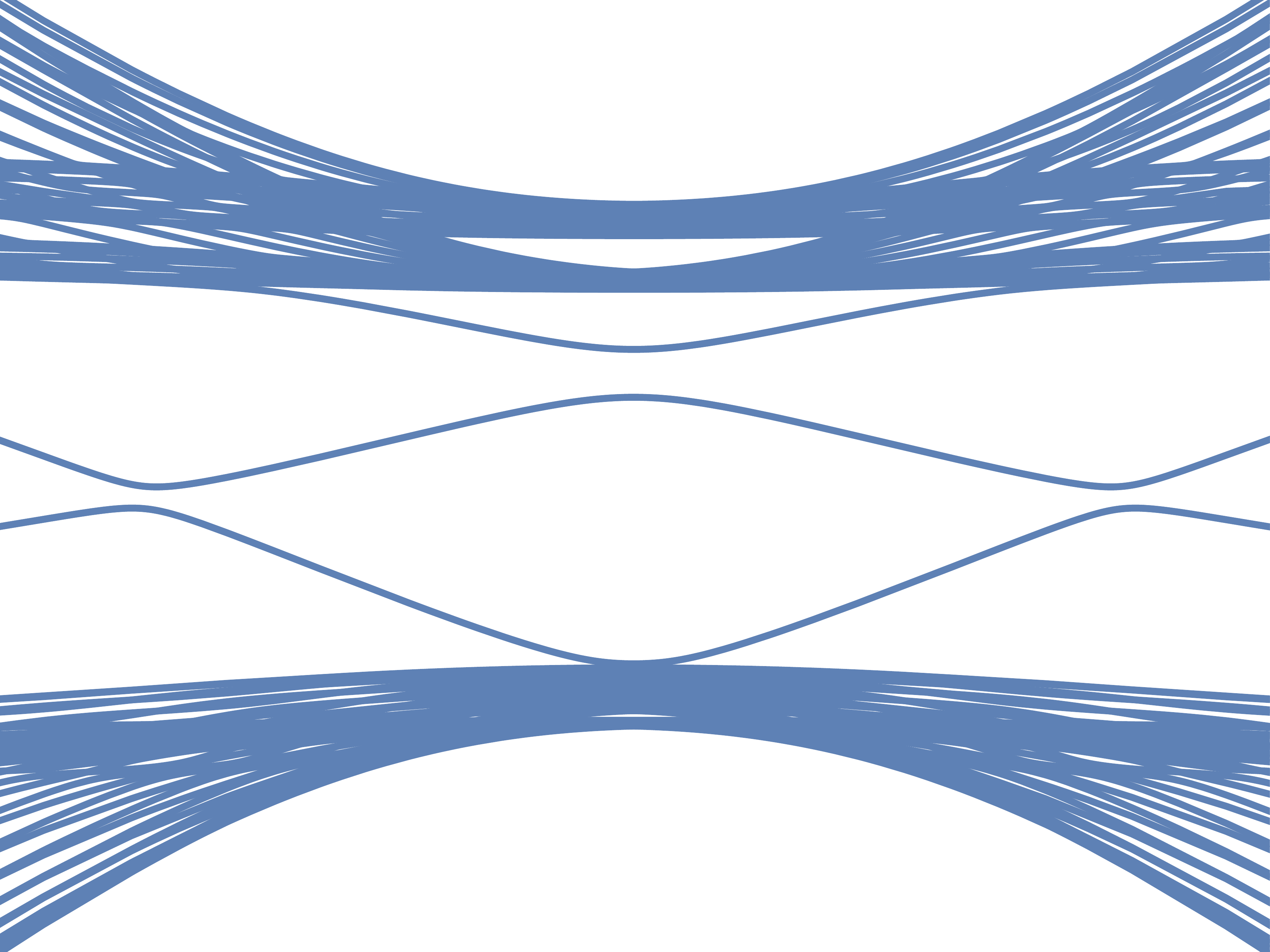}
\end{center}
\caption{ Phonon edge modes. Top left: $a_1=-1$,  $b_1=0.05$, $d_2=0.2$, $a_2=a_3=b_2=b_3=0$ (Phase Ib). Top right: $a_1=-1$, $a_3= -0.6$, $b_1=0.05$, $d_2=0.2$, $a_2=b_2=b_3=0$ (Phase IIb). Botton: close-ups of the gaps. }
\label{fig:f4}
\end{figure}

In order to compute the edge modes we have imposed periodic boundary conditions in the $x$-direction (along $\mathbf{e_1}$ in Fig~\ref{fig:f1}) and open boundary conditions on the two zig-zag edges. The results are presented in  Fig~\ref{fig:f4}, where $k_x$ goes from $0$ to $2\pi$. The edge spectrum in phase Ib (IIb) is qualitatively identical to the spectrum in phase Ia (IIa). The edge modes present several peculiarities. First of all, besides the usual gap-crossing `optical' edge modes, there are also `acoustic' edge modes below the bulk acoustic branch.  Secondly, the edge mode spectrum is gapped, as shown in the close-ups in 
Fig~\ref{fig:f4}. This is typical of topologically fragile phases~\cite{PhysRevLett.121.126402}, and is consistent with the negative coefficients in the linear combinations of EBRs in  Table~\ref{table:t2}. 

Note that if we set $b_1=b_2=b_3=0$  in Eq.~(\ref{dynmat})  the dynamical matrix becomes diagonal in `spin' space, i.e., right and left-handed modes decouple form each other. In that case, the edge modes become helical modes and the gaps in their spectrum disappear. As the $b_i$ bulk couplings are allowed by all the symmetries in the system, this is a sign of fragile topology. In the gapless limit $b_i=0$, acoustic and optical edge modes have opposite helicities: if the acoustic right moving edge modes have right circular polarization,  then then optical right moving edge modes have left circular polarization and vice versa.
Notice also that the edge mode spectrum is more complicated in the IIa and IIb phases, possibly reflecting the winding number two in the Wilson line spectrum. 

\begin{figure}[h]
\begin{center}
\includegraphics[angle=0,width=.45\linewidth]{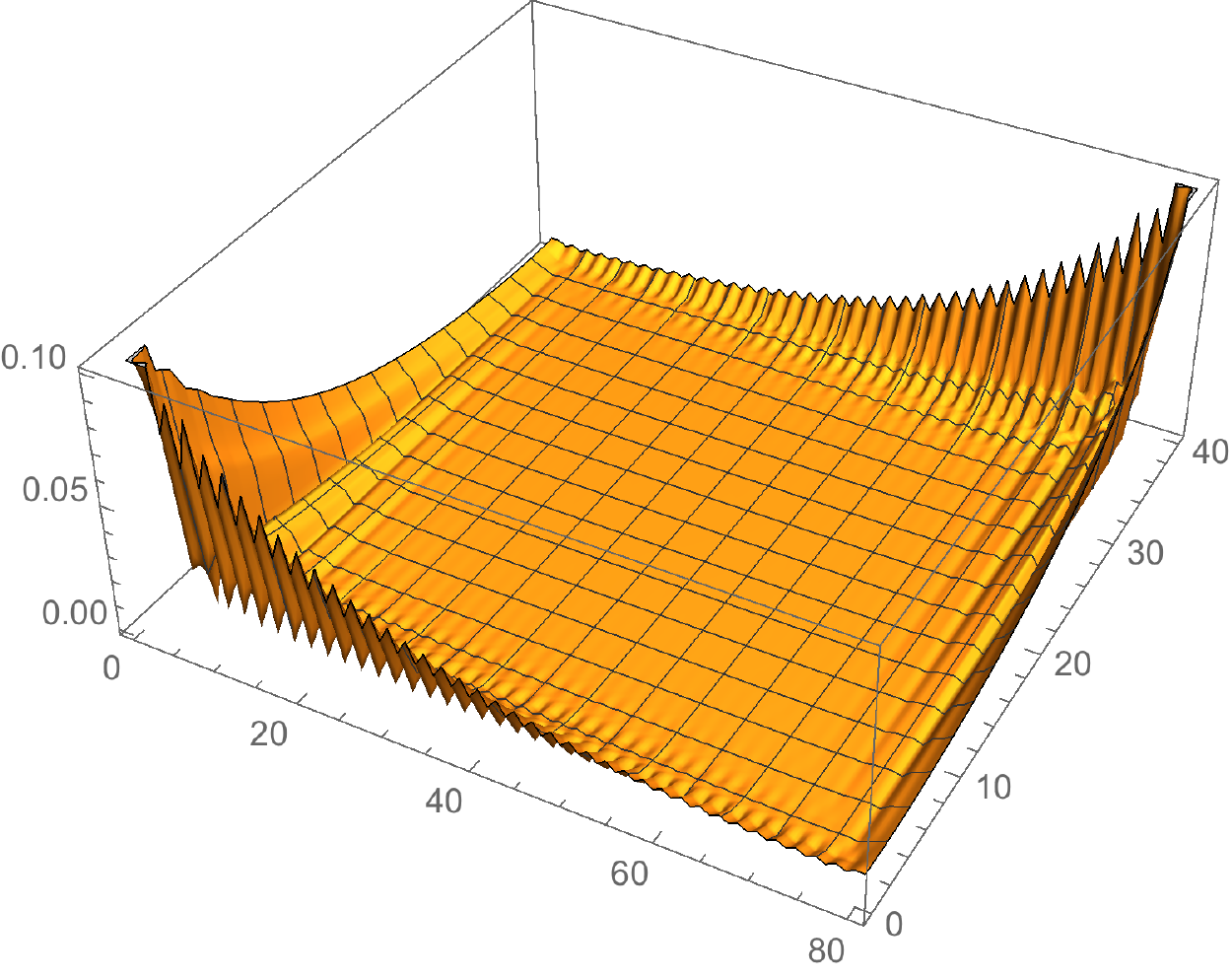}
\includegraphics[angle=0,width=.45\linewidth]{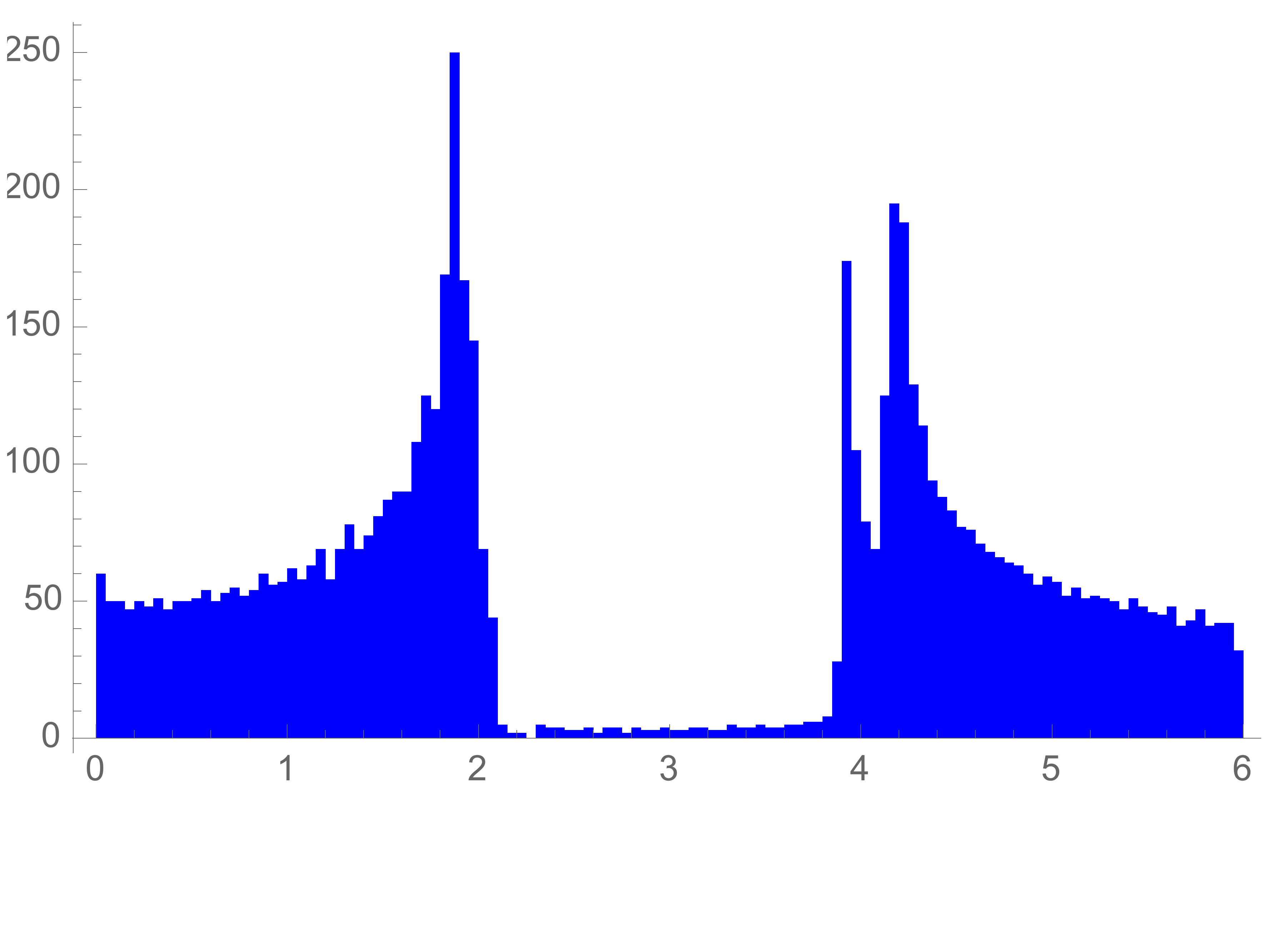}
\includegraphics[angle=0,width=.35\linewidth]{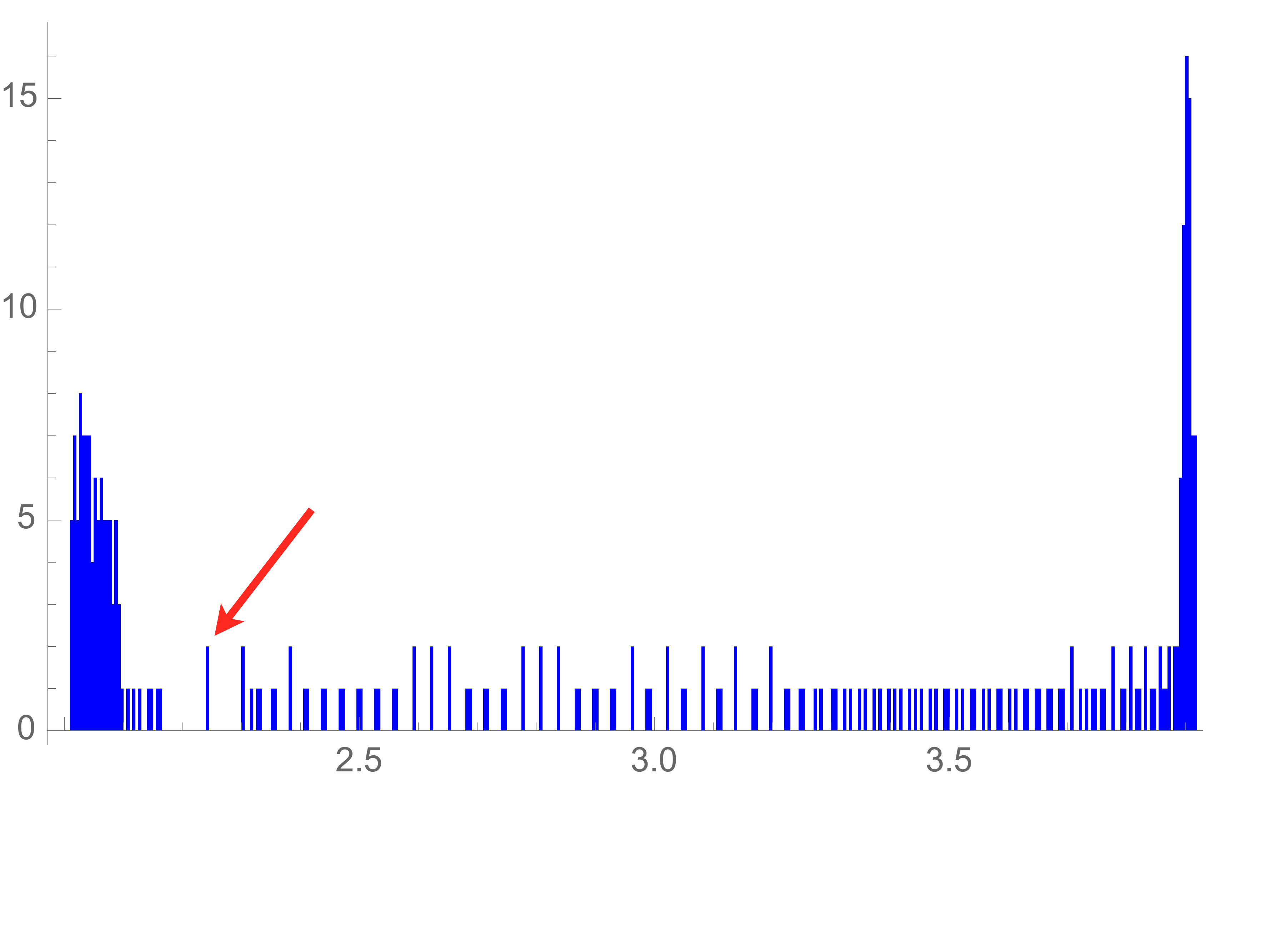}
\end{center}
\vspace*{-.6cm}
\caption{ Phononic corner   modes for \hbox{$a_1=-1$},  $b_1=0.05$, $d_2=0.2$, $a_2=a_3=b_2=b_3=0$ (Phase~Ib). Top left: Oscillation amplitudes for the  corner modes. Top right: Phonon DOS. Botton: DOS of the bulk gap, with arrowed degenerate corner modes. }
\label{fig:f5}
\end{figure}

Lastly, there are also corner modes, as shown in Fig~\ref{fig:f5}. This is again to be expected in topologically fragile phases, and is  considered a signature of higher order topology~\cite{2017Sci...357...61B,PhysRevB.96.245115,2018arXiv180611116W,2018arXiv181002373W}. 
The corner modes in Fig~\ref{fig:f5} have been computed for a parallelogram cut along the unit vectors in Fig~\ref{fig:f1}, with $40$ unit cells per side and open boundary conditions along the resulting zig-zag edges. The corner modes appear at the $60$ degree angles. For different values of the parameters we have also observed additional corner modes at the $120$ degree angles, sometimes buried in the continuum. The slow damping rate of the corner modes along the edges is a reflection of  the smallness of the gap in the edge mode spectrum. Corner modes are present in all four topological phases, but they are easier to observe in phases Ia and Ib, i.e., in those with winding number one.  

Concerning the topological stability of the corner modes, please see the analysis in Appendix~\ref{sec:fill}  where an attempt is made to relate the observed corner states  to  the recently proposed filling anomaly~\cite{PhysRevB.99.245151}. Although the existence of non-vanishing secondary topological indices for this system suggests that at least some of the numerically computed corner states must  be topologically protected, our analysis is inconclusive as to which ones are topologically robust. 

\section{Discussion}\label{discussion}
 Topological phononics is a very young field going back to less than a decade ago. In spite of the activity displayed, of which Refs.~\cite{Wang_2015,Kariyado:2015kq,PhysRevB.96.064106,PhysRevLett.115.104302,PhysRevLett.119.255901,2015Sci...349...47S,2016PNAS..113E4767S,Huber:2016fk,Kane:2013uq,PhysRevLett.117.068001,PhysRevB.86.035141,PhysRevLett.120.016401,PhysRevB.97.054305,PhysRevB.97.155422,PhysRevMaterials.2.114204} are but a small sample, only a handful of systems with topological phonon spectra are presently known, most of them mechanical contraptions, and virtually no material realizations have been described.  The results reported in this paper show that the methods of TQC can be adapted to a systematic search for topologically non-trivial phonons, with the crystal structure as input data. As these methods are susceptible of automatic implementation, this may eventually enlarge enormously our knowledge of materials which are the phononic analogues of topological insulators.~\cite{foots8}

One new insight from the analysis in this paper is that any material with the structure of the honeycomb has the potential to host topological phonons while keeping all its symmetries intact. This is, for phonons, the analogue of the Kane-Mele mechanism for electrons~\cite{PhysRevLett.95.226801,PhysRevLett.95.146802}. Indeed, the dynamical matrix studied in this paper can be considered as the fragile topology version of the Kane-Mele model. This should be contrasted with other approaches in the literature, where one tries to open a gap in the phonon spectrum by breaking time reversal symmetry, either through  Coriolis forces~\cite{Wang_2015,Kariyado:2015kq,PhysRevB.96.064106} or by interactions with external magnetic fields~\cite{Holz:1972fk,PhysRevLett.95.155901,PhysRevLett.96.155901}.

There are some open questions for the future. One  is the precise definition of fragile topology for phonons. When the irreps of an isolated subset of bands can be obtained as a difference $BR_1-BR_2$ of band representations, the addition of a trivial band that transforms according to $BR_2$ `trivializes' the fragile topology~\cite{PhysRevLett.121.126402}. In the case of electrons, the states transforming as $BR_2$ can, in principle, be found in the crystal, maybe as core orbitals deep in the valence bands, maybe forming high energy conduction bands. Obviously this is not the case for phonons, where the number of bands is limited by the crystal structure. Another question is  the possible connectivity constraints derived from the existence of acoustic bands. This problem does not arise  in graphene, where  in- and off-plane modes decouple, but may be an issue with other systems.

\begin{acknowledgments} 
%\paragraph{Acknowledgments}
It is a pleasure to thank A. Bernevig, B. Bradlyn, J. Cano and Z. Song for useful discussions and suggestions. This work has been supported in part  by Spanish Science Ministry grant PGC2018-094626-B-C21 (MCIU/AEI/FEDER, EU)  
and by Basque Government grant IT979-16.

\end{acknowledgments}

\appendix

\section{Parametrization of the dynamical matrix}
\label{pardyn}
The symmetry constraints on the  matrix of force constants for the planar honeycomb have been considered in Ref.~\onlinecite{FALKOVSKY20085189}. For the sake a completeness and to set the notation we give here a brief summary of the analysis for the in-plane modes that, according to the analysis in Section~\ref{sec:bandrep}, are the ones that can exhibit nontrivial topology. 
%The dynamical matrix $D(\mathbf{k})$ is the Fourier transform of 
The harmonic potential energy for the in-plane modes is given by 
\beq
U=\frac{1}{2}\sum_{i,j}  \mathbf{r}_i^t   \mathrm{U}_{ij}\mathbf{r}_j,
\eeq
 where $i,j$ run over all the atoms in the lattice, \hbox{$\mathbf{r}_i=(x_i,y_i)$} is the displacement of atom~$i$ from  equilibrium  and $U_{ij}=U_{ji}^t$ is the matrix of   force constants between atoms $i$ and $j$. 

Assuming  $i,j$ are nth-nearest neighbors, the $2\times 2$ matrix $ \mathrm{U}_{ij}$ is  parametrized by four real coefficients $(a_n,b_n,c_n,d_n)$. Pick one nth-nearest neighbor $j$ to the atom $i$ and parametrize $U_{ij}$ as
\beq
U_{ij}=\left(\begin{array}{cc}
a_n+b_n & -c_n-d_n\\
-c_n+d_n & a_n-b_n\\
\end{array}
\right),
\eeq
where the parameters are real by time reversal symmetry. Then, if $(i'j')$ is another pair of nth-nearest neighbors such that $\mathbf{d}_{i'j'}=V(g)\mathbf{d}_{ij}$, where $V(g)$ is the vector representation for the symmetry operation $g$ that transforms $ij$ into $i'j'$, we will have
\beq\label{eq:trans}
U_{i'j'}=V(g) U_{ij} V(g)^{-1}.
\eeq

\begin{table}[h]
\begin{tabular}{| c | c | c | c | c | c |c |}
\hline
1 & 2 & 3 & 4 & 5 & 6 & 7 \\
\hline\hline
 $a_1$ & $a_2$ & $a_3$ & $a_4$ & $a_5$ & $a_6$ & $a_7$ \\
\hline
$b_1$ & $b_2$ & $b_3$ & $b_4$ & $b_5$ & $b_6$ & $b_7$ \\
\hline
$-$ & $-$ & $-$ & $c_4$ & $-$ & $-$ & $c_7$ \\
$-$ & $d_2$ & $-$ & $d_4$ & $-$ & $d_6$ & $d_7$ \\
\hline
\end{tabular}
\caption{Non-vanishing real coefficients parametrizing the harmonic potential between $nth$-nearest neighbors for \hbox{$n=1,\ldots,7$.} }
\label{table:ts1}
\end{table}

%Given that $6mm$ contains only $12$ elements, 
If the number of nth-nearest neighbors is sufficiently large, we may have several subsets of pairs $(ij)$ unrelated by symmetry. In that case we would need more that one set of parameters $(a_n,b_n,c_n,d_n)$. For the honeycomb lattice this will happen whenever the number of nth-nearest neighbors is greater than $12$, which is the number of elements in $6mm$. On the other hand, the number of independent parameters for given $n$ may be less that four. This happens  if the two atoms $(i,j)$ are  left invariant (exchanged) by a symmetry element $g$, since then Eq.~(\ref{eq:trans}) (combined with $U_{ij}=U_{ji}^t$) becomes a constraint. For the honeycomb lattice, one possibility is to have a mirror $m_\parallel$ along the link $\mathbf{d}_{ij}$. Assuming for simplicity that $\mathbf{d}_{12}=d_{12}\mathbf{\hat x}$,  under the action of $m_\parallel$  $(x_i,y_i)\to(x_i,-y_i)$ for $ i=1,2$. This implies $c_n=d_n=0$. Another possibility is to have a mirror $m_\perp$ perpendicular  to $\mathbf{d}_{ij}$ through the midpoint of the link. This exchanges the two atoms and reverses the $x$ components of the displacements, which implies $c_n=0$. The effect of these constraints is summarized in Table~\ref{table:ts1} for $n$ up to seven. It is important to realize that, in order to impose a constraint, the parallel mirror  $m_\parallel$ has to pass \textit{through} the two atoms. A mirror plane that is parallel to $\mathbf{d}_{ij}$ but does not go through $ij$ does \textit{not} impose any constraint on $U_{ij}$; it merely relates $U_{i'j'}$ to $U_{ij}$, where $(i'j')$ are the images of $(ij)$ by the mirror plane. Failing to appreciate this point would lead us to set $d_2=0$, as in Ref.~\onlinecite{FALKOVSKY20085189}.

These are the only possible constraints on the planar honeycomb lattice for $i\neq j$. For $i=j$, the $C_{3z}$ invariance about each atom implies $U_{ii}=a_0\openone$. Note, however, that $a_0$ is not an independent parameter, due to the global translation invariance of the crystal, which implies  $\sum_j U_{ij}=0$. This guarantees that atom $i$ does not experience any force when all the atoms in the crystal are given a uniform displacement $\mathbf{r}_i=\mathbf{t}$ and is behind the existence of  acoustic phonon branches. This is a peculiarity of phonon dynamics  without analog in  electron hamiltonians, where on-site energies are independent parameters. For the planar honeycomb lattice, as a the result of global translation invariance, we have  $a_0=-3a_1-6a_2-3a_3-\ldots$

\section{Symmetries of the dynamical matrix}\label{sec:symdyn}
The dynamical matrix is invariant under all the symmetries of the time-reversal symmetric honeycomb. The time-reversal operation may be written as $\mathcal{T}=\theta \mathcal{K}$, where $\mathcal{K}$ denotes complex conjugation and the unitary matrix $\theta$ is given by 
\beq
\theta=\openone_\sigma\otimes s_x\, ,
\eeq
reflecting the fact that time-reversal exchanges the two circular polarizations but not the sublattices.
Invariance of the dynamical matrix under time-reversal requires
\beq
\theta D(\mathbf{k})^*\theta^\dagger= D(-\mathbf{k}),
\eeq
since $\mathcal{T}$ reverses the sign of the momenta. 

By construction, the dynamical matrix $D(\mathbf{k})$ is periodic in reciprocal space, $D(\mathbf{k}+\mathbf{K})=D(\mathbf{k})$, where $\mathbf{K}$
is any vector in the reciprocal lattice. This follows trivialy from the definition~(\ref{ft}). In order to consider the invariance of the dynamical matrix under point group operations, it is convenient to consider the alternative definition
\beq\label{aft}
\tilde D_{ab}(\mathbf{k})=\sum_{\mathbf{R}}U_{a,\mathbf{0};b,\mathbf{R}}\,e^{i\mathbf{k}\cdot(\mathbf{R}-\boldsymbol{\delta}_b -\boldsymbol{\delta}_a )} \,,
\eeq
which is \textit{not} periodic but is related to $D(\mathbf{k})$ by a unitary transformation
\beq
\tilde D(\mathbf{k})=V^\dagger(\mathbf{k}) D(\mathbf{k}) V(\mathbf{k})\,,
\eeq
where
\beq
 V(\mathbf{k})=\left(\begin{array}{cc}
e^{i \mathbf{k}\cdot\boldsymbol{\delta}_A}\openone_s &0\\
0 &e^{i \mathbf{k}\cdot\boldsymbol{\delta}_B} \openone_s\\
\end{array}\right)\,.
\eeq
Then, invariance under a point group operation $R$ requires
\beq\label{invR}
M(R) \tilde D(\mathbf{k})M(R)^\dagger= \tilde D(R\mathbf{k}),
\eeq
where $M(R)$ denotes the corresponding matrix in the mechanical representation. In-plane displacements of the atoms are invariant under the horizontal plane and for all practical purposes we can use  the layer group $p6mm$. This is an infinite group that includes all the lattice translations, but we can restrict ourselves to the point group elements, as translation invariance has been implemented by taking the Fourier transform in Eq.~(\ref{ft}). The point group $6mm$ can be generated by the sixfold counterclockwise rotation $C_{6z}^+$ and a vertical plane $m_x$ perpendicular to the $OX$ axis. The mechanical representation for these generators is given by
\beq
M(C_{6z}^+)=\sigma_x\otimes \Gamma_6(C_{6z}^+), \, M(m_x)=\openone_\sigma\otimes \Gamma_6(m_x),
\eeq
where we have used the fact  that in-plane displacements transform according to the $\Gamma_6$ irrep. The presence of  $\sigma_x$ and $\openone_\sigma$ in the equation above reflects the fact that the sublattices are exchanged by $C_{6z}^+$ but not by $m_x$. The $\Gamma_6$ matrices for the generators are given by 
\beq
\Gamma_6(C_{6z}^+)=\left( \begin{array}{cc}
e^{i\pi/3} & 0 \\
0 & e^{-i\pi/3}\\
\end{array} \right)\!,
\Gamma_6(m_x)=\left( \begin{array}{cc}
0 & -1 \\
-1 & 0\\
\end{array} \right).
\eeq
Imposing Eq.~(\ref{invR}) for the two generators of the point group $6mm$ guarantees the invariance of the dynamical matrix under the whole layer group $p6mm$.

\section{Constraints on Wilson loop winding}\label{sec:wlw}
\subsection{$C_{2z}$-eigenvalues and WL winding}
We mentioned in Section~{\ref{sec:wlwsym} that the winding of the WLs can be understood in terms of the $C_{2z}$-eigenvalues of the normal modes at the two $C_{2z}$-invariant points $\Gamma$ and $M$. Here we fill in the details of the argument following the ideas in Ref.~\onlinecite{PhysRevB.89.155114} and  taking into account that,  for the in-plane modes of the honeycomb lattice, $C_{2z}$ plays the role of  inversion symmetry. The $\mathbf{g_1}$-directed WL is defined by
\beq
 W(k_2) \equiv  P e^{i\int_{0}^{2\pi} dk_1 A_1(k_1,k_2)}.  
\label{eq:WL}
\eeq
For fixed $k_2$, this formula can be interpreted as the WL for a 1-dimensional system   with reciprocal  primitive cell  $k_1\in [0,2\pi]$ along $\mathbf{g_1}$. Then, for values of $k_2$ such that the 1-dimensional system is $C_{2z}$-invariant, i.e., for $k_2=0,\pi$, the number $N_{(-1)}$ of $-1$ eigenvalues of the WL is given by Eq.~(1) in Ref.~\onlinecite{PhysRevB.89.155114}, namely
\beq
N_{(-1)}=|n_{(-)}(0)-n_{(-)}(\pi)|,
\eeq
where $n_{(\pm)}(0)$ and $n_{(\pm)}(\pi)$ are the numbers of normal modes with $C_{2z}$ eigenvalue equal to $\pm 1$ at $k_1=0$ and $k_1=\pi$ respectively. Similarly, the number of complex conjugate pairs of eigenvalues $(\lambda,\lambda^*)$ of the WL is given by~\cite{PhysRevB.89.155114}
\beq\label{ns}
n_s=\mathrm{Min}\{  n_{(\pm)}(0),n_{(\pm)}(\pi) \}.
\eeq

Now, as Fig.~\ref{fs2} shows, for $k_2=0$ the 1-dimensional WL goes through $\Gamma$ and $M''$, and we have
\beq
N_{(-1)}(k_2=0)=|n_{(-)}(\Gamma)-n_{(-)}(M'')|.
\eeq
This is equal to $2$ for phases Ia and Ib, where we have two modes with   $C_{2z}$-eigenvalues $\eta$ at  $\Gamma$ and $-\eta$ at $M$, and to~$0$ for phases IIa and IIb, where the eigenvalues are equal to $\eta$ at  $\Gamma$ \textit{and} $M$. Note that $M''$ and $M$ are related by a unitary  $C_{3z}$-rotation that does not change the $C_{2z}$ spectrum. On the other hand, by Eq.~(\ref{ns}) the number of complex conjugate pairs of eigenvalues $(\lambda,\lambda^*)$ of the WL is zero in all cases. 
As the WL eigenvalues are equal to $e^{2\pi i x_1(k_2)}$, where $x_1(k_2)$ is the hybrid Wannier function center along $\mathbf{e}_1$, this implies $x_1(\Gamma)= 1/2$ for phases Ia and Ib and 
$x_1(\Gamma)=0$ for phases IIa and IIb. This is exactly what is observed in Fig.~\ref{fig:f3}. 

For $k_2=\pi$ the 1-dimensional WL goes through $M$ and $M'$, and we have
\beq
N_{(-1)}(k_2=\pi)=|n_{(-)}(M)-n_{(-)}(M')|=0,
\eeq
irrespective of the topological phase. As above, Eq.~(\ref{ns}) implies that the number of complex conjugate pairs of eigenvalues $(\lambda,\lambda^*)$ of the WL is zero in all cases. Thus $x_1(M)= 0$ for all the topological phases, in agreement with Fig.~\ref{fig:f3}.
\begin{figure}[h]
\begin{center}
%\vspace*{-.3cm}
\includegraphics[angle=0,width=1.\linewidth]{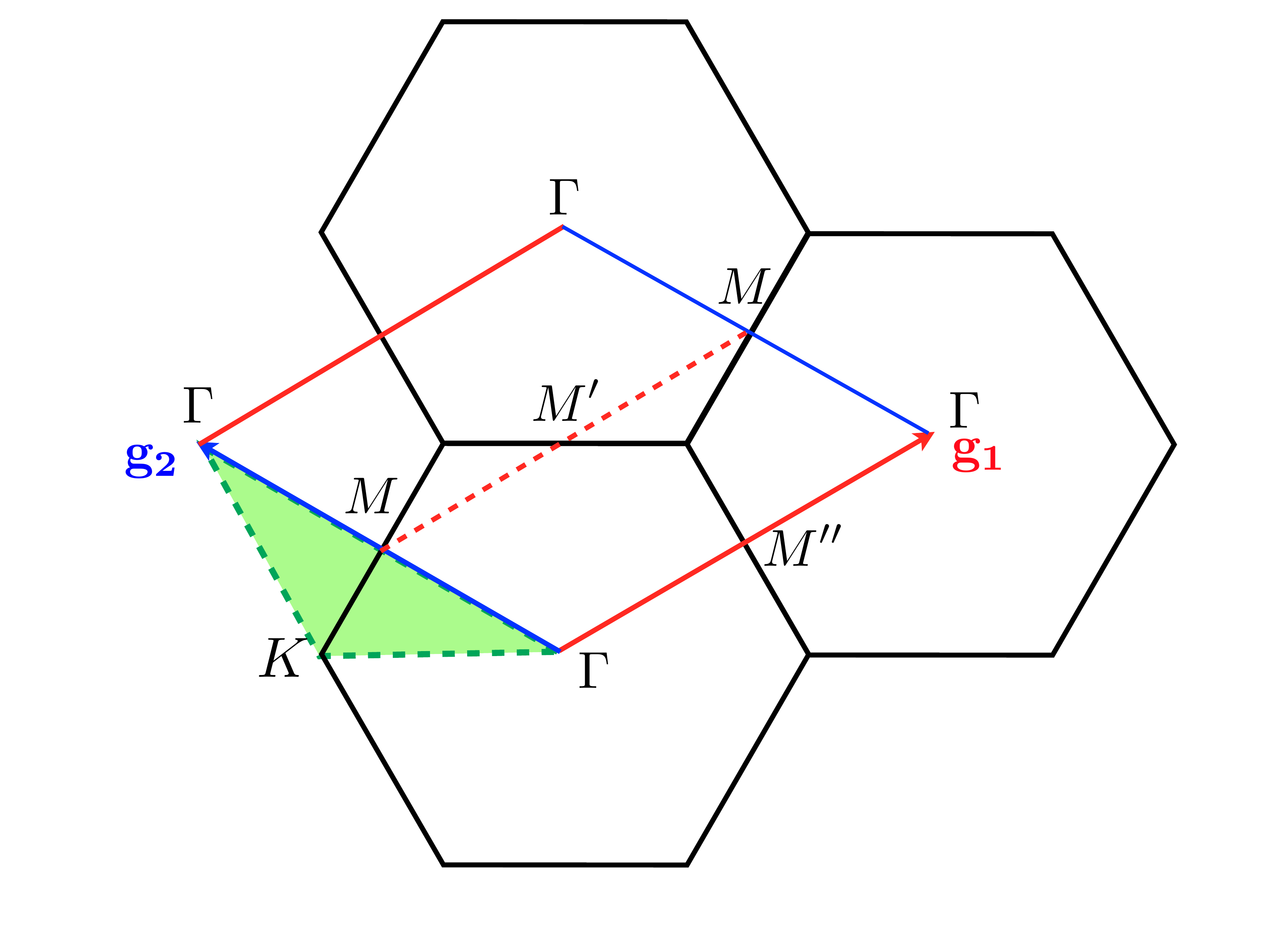}
\end{center}
\vspace*{-.3cm}
\caption{One-dimensional Wilson loops for $k_2=0$ (red full   line) and $k_2=\pi$ (red dashed line). The green dashed contour is used in subsection~\ref{sc} to extract the consequences of \hbox{$C_{3z}$-invariance}.}
\label{fs2}
\end{figure}
For phases IIa and IIb the hybrid Wannier functions sit at the same point for $k_2=0$ and $k_2=\pi$, and we could have non-winding WLs. This is obviously impossible for 
phases Ia and Ib, where the WLs necessarily must wind. 

\subsection{Crossings at generic points protected by $C_{2z}\mathcal{T}$}
Having the same hybrid Wannier centers at $k_2=0$ and $k_2=\pi$ is compatible with non-winding WLs, but also with even winding number. For phases IIa and IIb we find the the WLs have winding number two for all the disconnected branches. As shown in Fig.~\ref{fig:f3}, this involves crossings at generic values of $k_2$, and we might worry at their stability against small perturbations. Here we present for phonons the analog of  the arguments given in Refs.~\onlinecite{2018arXiv180409719B,PhysRevB.99.045140,2018arXiv180710676S} for the stability of generic crossings for electrons. Note that for spin-$1/2$ systems \hbox{$C_{2z}^2\!=\!\mathcal{T}^2\!=\!-1$}, whereas for bosons we have \hbox{$C_{2z}^2=\mathcal{T}^2=1$}.

The combined operation $C_{2z}\mathcal{T}$ leaves $(k_1,k_2)$ invariant, and the WL loop must satisfy the constraint
\beq\label{c2t}
C_{2z}\mathcal{T}W(k_2) (C_{2z}\mathcal{T})^{-1} = W(k_2).
\eeq
Writing the unitary operator $W(k_2)$ in terms of the `Wannier hamiltonian' $H_W$
\beq
W(k_2)=e^{i H_W(k_2)},
\eeq
the constraint becomes
\beq\label{const}
C_{2z}\mathcal{T}H_W(k_2) (C_{2z}\mathcal{T})^{-1} = -H_W(k_2),
\eeq
where the extra minus sign is due to the fact that $C_{2z}\mathcal{T}$ is an antiunitary operation 
\beq
C_{2z}\mathcal{T}=U\mathcal{K},
\eeq
where $U$ is a unitary matrix and $\mathcal{K}$ denotes complex conjugation. In order to determine the form of the matrix $U$ we  note that, according to Eq.~(\ref{c2t}), for generic $k_2$ the WL is invariant under the magnetic group $2'$ that has only two elements, $2'=\{E,C_{2z}\mathcal{T}\}$, where $E$ is the identity operation. Now, according to Ref.~\onlinecite{Bradley}, the only single-valued irreducible corepresentation for the group $2'$ is 1-dimensional, with the unitary matrix for $C_{2z}\mathcal{T}$ given by $D(C_{2z}\mathcal{T})=\pm 1$. The two signs give unitarily equivalent corepresentations, and we may take the $+$ sign without loss of generality. As we are considering two-band WLs, these must transform as the 2-dimensional corepresentation obtained by taking two copies of the 1-dimensional irreducible  corepresentation. This means that $U=\openone_2$ and Eq.~(\ref{const}) reduces to
\beq\label{const1}
\mathcal{K}H_W(k_2)\mathcal{K}=H_W(k_2)^*=-H_W(k_2).
\eeq
Writing $H_W$ as a linear combination of Pauli matrices
\beq
H_W(k_2)=a_0(k_2)\openone_2+a_x(k_2)\sigma_x+a_y(k_2)\sigma_y+a_z(k_2)\sigma_z,
\eeq
we see that the constraint~(\ref{const1}) implies
\beq\label{sol}
H_W(k_2)=a(k_2)\sigma_z.
\eeq 
Had we taken two copies of the 1-dimensional irreducible representation with different signs for $D(C_{2z})$, the constraint would read
\beq
\sigma_z H_W(k_2)^* \sigma_z=-H_W(k_2),
\eeq
with  solution
\beq
H_W(k_2)=a(k_2)\sigma_x,
\eeq 
that is related to (\ref{sol}) by a unitary transformation. In any case, due the periodicity of the Wannier hamiltonian eigenvalues, there will be a crossing whenever $a(k_2)=n\pi$, and this may happen for generic values of $k_2$. As we can not add another Pauli matrix, small perturbations will merely shift the position of the crossings, but will not be able to remove them. Note that $a(k_2)=n\pi$ corresponds to $x_1(k_2)=n/2$, i.e., the protected crossings will take place for $x_1=0, 1/2$. Thus the generic crossings at $x_1=1/2$ in Fig~\ref{fig:f3} are protected by $C_{2z}\mathcal{T}$ and the winding of the WL is  the winding of the function $a(k_2)$.

\begin{figure}[h]
\begin{center}
\includegraphics[angle=0,width=1.\linewidth]{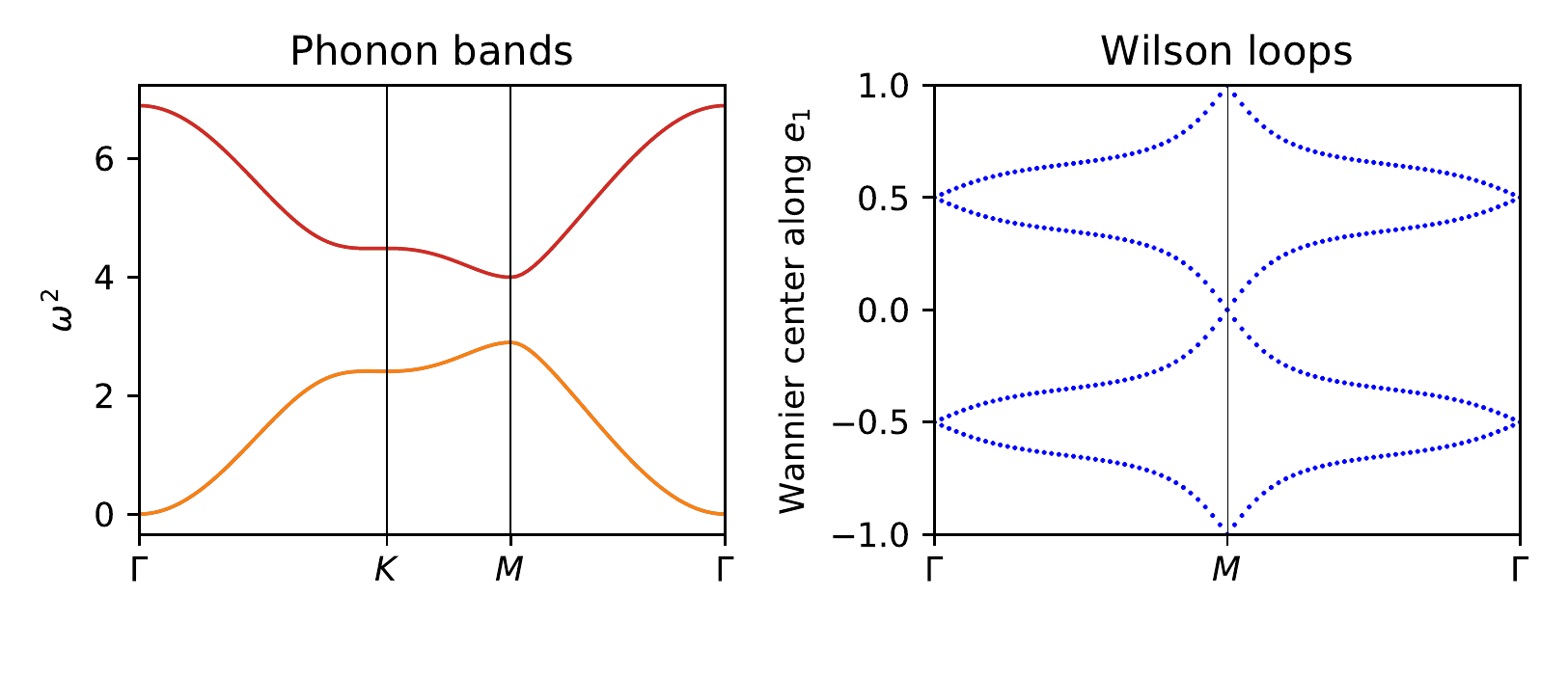}
\includegraphics[angle=0,width=1.\linewidth]{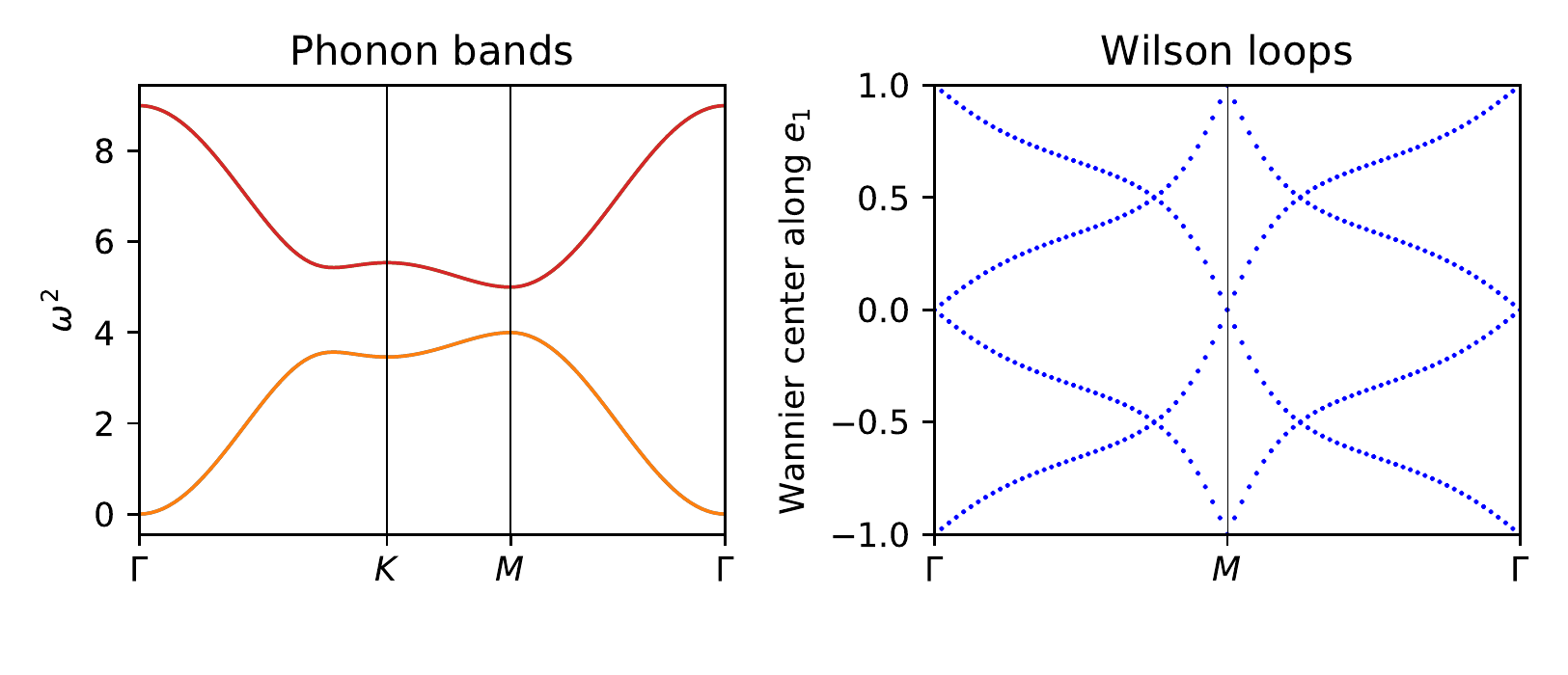}
\end{center}
\vspace*{-.8cm}
\caption{Phonon bands (left)  and Wannier centers for the acoustic bands (right). Top: $a_1=-1$, $a_3= -0.15$,  $d_2=0.2$, $a_2=b_1=b_2=b_3=0$ (Phase Ib). Bottom: $a_1=-1$, $a_3= -0.5$,  $d_2=0.2$, $a_2=b_1=b_2=b_3=0$ (Phase IIb) }
\label{fs3}
\end{figure}

\subsection{The role of $C_{3z}$-invariance}\label{sc}
As shown in Refs.~\onlinecite{2018arXiv180409719B,PhysRevB.99.045140,2018arXiv180710676S}, in order to extract the consequences of $C_{3z}$-invariance on WL winding, one has to choose the paths of integration with some care. In this subsection we  apply  to our system the results in Ref.~\onlinecite{2018arXiv180409719B}, to which we refer the reader for a detailed presentation.  There it is shown that the WL for a single isolated band along the green dashed  path in Fig.~\ref{fs2}  is given by 
\beq\label{w3}
e^{i\gamma}=\xi_2^\Gamma \xi_3^K(\xi_2^M\xi_3^\Gamma)^{-1},
\eeq
where  $\xi_n^\mathbf{k}$ is the $C_{nz}$-eigenvalue of the normal mode at the high symmetry point $\mathbf{k}$. As the loop encloses one sixth of the BZ, one can use the symmetries of the system~\cite{2018arXiv180409719B} to show that the total WL winding number for the single isolated band is given by 
\beq\label{gamma}
w=6 \gamma/2\pi=3\gamma/\pi.
\eeq

It is important to note that, in order to obtain this result, one has to use other symmetries of the system besides $C_{3z}$. In their absence, as emphasized in Ref.~\onlinecite{PhysRevB.99.045140}, $C_{3z}$-invariance alone may not be enough to guarantee the existence of nontrivial windings for a particular WL, even though nontrivial windings will show up in other, appropriately chosen WLs.

In order to apply Eq.~(\ref{w3}) to our system, where we have isolated  \textit{pairs}  of bands, we turn off all the $b_i$ couplings. As mentioned in Section~\ref{edgecorner} , this effectively decouples the right- and left-polarized sectors, and we end up with two 2-band systems with identical frequency spectra. This is shown in Fig.~\ref{fs3}, which should be compared with Fig.~\ref{fig:f3}. Note, in particular, that setting $b_1=0$ does not change the WL winding numbers. That is to be expected, as we can vary the values of  the $b_i$ couplings adiabatically without closing the global gap that separates  the acoustic and optical branches. This means that we can use  Eq.~(\ref{w3}) to compute the winding number for the isolated bands in the decoupled left and right-polarized systems, knowing that the results are valid also for $b_i\neq 0$. See Ref.~\onlinecite{2018arXiv180409719B} for a thorough discussion of this point and further examples with spinful electrons.

As the dynamical matrix can be diagonalized analytically at the three high symmetry points, it is straightforward to compute the required $C_{nz}$-eigenvalues. We find that the WL for phases Ia and Ib is given by 
\beq
e^{i\gamma}=e^{\pm \frac{i\pi}{3}},
\eeq
which implies $\gamma=\pm \pi/3 +2\pi n$, where $n\in \mathbb Z$. By Eq.~(\ref{gamma}), this translates into 
\beq\label{wI}
w_I=6n\pm 1.
\eeq
The windings in Fig.~\ref{fs3} (top) are obtained by setting \hbox{$n=0$}. Similarly, the result for phases IIa and IIb is
\beq
e^{i\gamma}=e^{\pm \frac{i2\pi}{3}},
\eeq
which implies $\gamma=\pm 2\pi/3 +2\pi n$, yielding 
\beq\label{wII}
w_{II}=6n\pm 2,
\eeq
where the windings in Fig.~\ref{fs3} (bottom) are again obtained by setting $n=0$. 
Combining  (\ref{wI}) and (\ref{wII}) gives $w=3n \pm 1$, which is the result quoted at the end of  Section~\ref{sec:wlwsym}. Thus zero winding number is incompatible with $C_{3z}$-invariance and  any isolated branch, acoustic or optical,   must be topologically nontrivial.

\section{Corner modes and filling anomaly}\label{sec:fill}

According to Ref.~\onlinecite{PhysRevB.99.245151}, corner-localized quantized features are protected by secondary topological indices that can be computed from the symmetry eigenvalues at the high symmetry points of the BZ of systems with \hbox{$C_n$-symmetry.} Those indices are a manifestation of the filling anomaly that, for electron bands, can be understood as a mismatch between the number of electrons  required for charge neutrality and to preserve the crystal symmetry.
This concept has been recently applied to other types of bands, such as those in photonic~\cite{PhysRevA.101.043833,he2019quadrupole} and acoustic crystals.~\cite{lin2020higherorder,peri2019experimental}

\begin{widetext}{\ }
\begin{table}[h]
\renewcommand{\arraystretch}{1.2}
\begin{tabular}{|c|c c |c c c c||c|c c |c c c |}
\hline
  $C_{2z}$ & $\Gamma_5$ & $\Gamma_6$ & $M_1$ & $M_2$& $M_3$ & $M_4$& $C_{3z}$ & $\Gamma_5$ & $\Gamma_6$ & $K_1$ & $K_2$& $K_3$  \\
\hline\hline
& $\left(\begin{array}{cc}
1 &0\\
0 &1\\
\end{array}\right)$ 
& $\left(\begin{array}{cc}
-1 &0\\
0 &-1\\
\end{array}\right)$ &1 & 1& -1&-1 
& &$\left(\begin{array}{cc}
e^{\frac{2 \pi i}{3}} &0\\
0 &e^{-\frac{2 \pi i}{3}}\\
\end{array}\right)$ &$\left(\begin{array}{cc}
e^{\frac{2 \pi i}{3}} &0\\
0 &e^{-\frac{2 \pi i}{3}}\\
\end{array}\right)$ &1 & 1&$\left(\begin{array}{cc}
e^{\frac{2 \pi i}{3}} &0\\
0 &e^{-\frac{2 \pi i}{3}}\\
\end{array}\right)$ \\
\hline
\end{tabular}
\caption{Matrices for the relevant irreps at the $\Gamma$, $M$ and $K$ points used to compute the topological index~$Q^{(6)}_{corner}$.}
\label{table:t7}
\end{table}

\end{widetext}

The appropriate secondary topological index for a  $C_6$-symmetric crystal is given by~\cite{PhysRevB.99.245151}
\beq\label{qcorner}
Q^{(6)}_{corner}=\frac{1}{4}[M_1^{(2)}]+\frac{1}{6}[K_1^{(3)}] \mod 1
\eeq
with
\beq\label{Pip}
[\Pi_p^{(n)}]\equiv \# \Pi_p^{(n)}-\# \Gamma_p^{(n)},
\eeq
where $\# \Pi_p^{(n)}$ counts the number of $e^{2\pi i(p-1)/n}$ eigenvalues of $C_n$ at the high symmetry point $\Pi$ and we take the $\Gamma$ point as a reference. The eigenvalues should be counted for all the bands below the gap or, more generally, for an isolated subset of bands.
Eqs.~(\ref{qcorner}) and~(\ref{Pip}) can be easily evaluated with the help of the matrices in Table~\ref{table:t7} and the irreps in Table~\ref{table:t1}. The results are summarized in Table~\ref{table:t8}. Note that, according to Table~\ref{table:t2}, the two subsets of bands with vanishing $Q^{(6)}_{corner}$ can be induced from a Wannier function at the center of the hexagon, wich is the Wigner-Seitz primitive cell that preserves the $C_6$-symmetry of the honeycomb lattice.

Table~\ref{table:t8}  shows that, for at least one subset of bands in each  phase,  the topological index $Q^{(6)}_{corner}$ takes  a non-vanishing value.   This suggests that  some of the observed corner modes must be  topologically protected, although pointing out which ones may require a more detailed analysis  that we leave for future work. Also for the future is left the task of clarifying the meaning of concepts such as charge neutrality and fractional charges in a phononic context. See Ref.~\onlinecite{miert2020topological} for a very recent  illustration of the subtleties involved in elucidating the topological character of corner modes.

\begin{table}[h]
\vspace*{.4cm}
\begin{tabular}{|c || c | | c |  }
\hline
Phase & Acoustic branch& Optical branch\\
\hline
\hline
Ia &  $5/6$ & $1/2$ \\
\hline
Ib &  $1/2$ & $5/6$ \\
\hline
IIa & $1/3$ & $0$ \\
\hline
IIb &  $0$ & $1/3$ \\
\hline \end{tabular}
\caption{Values of $Q^{(6)}_{corner}$ for the different isolated branches. }
\label{table:t8}
%\vspace*{-.3cm}
\end{table}

\bibliography{PhonTop_PRB}

\begin{thebibliography}{73}
\bibitem [{foot1()}]{foot1}%
  \BibitemOpen
  \bibinfo {note} {The index $i$ runs over all the atomic orbitals or, more
  generally, Wannier functions in a primitive unit cell. We can also consider a
  subset of orbitals, as long as they are closed under all the symmetries of
  the system.}\BibitemShut {Stop}%
\end{thebibliography}
\end{document}